# Subspace Identification of Large-Scale 1D Homogeneous Networks

Chengpu Yu, Michel Verhaegen, and Anders Hansson

***Abstract*—This paper considers the identification of large-scale 1D networks consisting of identical LTI dynamical systems. A new subspace identification method is developed that only uses local input-output information and does not rely on knowledge about the local state interaction. The identification of the local system matrices (up to a similarity transformation) is done via a low dimensional subspace retrieval step that enables the estimation of the Markov parameters of a locally lifted system. Using the estimated Markov parameters, the state-space realization of a single subsystem in the network is determined. The low dimensional subspace retrieval step exploits various key structural properties that are present in the data equation such as a low rank property and a *two-layer* Toeplitz structure in the data matrices constructed from products of the system matrices. For the estimation of the system matrices of a single subsystem, it is formulated as a structured low-rank matrix factorization problem. The effectiveness of the proposed identification method is demonstrated by a simulation example.**

## I. Introduction

In recent years, the fundamental system theoretical research on large-scale interconnected systems has been intensified both on the topic of distributed control, see e.g. [1]–[4], as well as on the topic of distributed system identification, see e.g. [5]–[7]. These fundamental developments are inspired by the increasing interest from the application side, such as in fluid mechanics [8], flexible structures [9], large adaptive telescope mirrors [10], wind turbine farms [11], and so on. In this paper, we focus on the identification of large-scale 1D homogeneous networked systems. Such systems may characterize a platoon of cars or may result by discretizing dynamical systems described via partial differential equations [3], [12], [13].

When considering the subspace identification of local system dynamics using local data only, there is not only the missing information about the local system state, but also the missing state information from the neighboring systems. This makes the local identification problem more difficult. In [14] a subspace identification (SID) scheme for large-scale circulant systems was developed. The specific property of circularity was exploited to decompose the large model description into simple subsystems through a global transformation of the input and output data. In [15] a network of identical systems interacting in a known interconnection pattern, indicated as so-called decomposable systems, was considered. The identification of the dynamics of a single subsystem in the network using transformed input-output data sequences gives rise to a challenging Bilinear Matrix Inequality (BMI). In [12] the property that the inverse of the observability Grammian is off-diagonally decaying was used to approximate the neighboring states, influencing the local system dynamics to be identified, via an (unknown) linear combination of locally neighboring inputs and outputs. The selection of these neighboring input and output quantities requires however an exhaustive search, which is quite computationally demanding. As a complementary work to [12], a nuclear norm identification solution was provided in [16] to separate the local dynamics and global dynamics by exploiting their distinct rank and order properties; however, it did not consider identification of the interconnections between the subsystems.

In this paper, we consider the subspace identification of a local cluster of identical LTI systems connected in a 1D network. Due to the unknown neighboring-state information, the considered local identification is a blind identification problem, where some input sequences are totally unknown. In this scenario, the existing SID approaches that aim at retrieving a matrix whose row or column space is of interest, the latter, e.g., being the extended observability matrix, cannot work. As a result, we aim to develop a new identification method for identifying the system matrices of a single subsystem (up to a similarity transformation) in the network that uses local input-output data only.

The subspace identification method presented in this paper has two major features. *First*, a novel identification method is developed that uses similar data equations as many existing SID variants, but aims for a low dimensional subspace retrieval that allows the identification of the finite number of Markov parameters present in this data equation instead. This is a new alternative to classical SID methods [17] that would fail in retrieving subspace like the column space of the extended observability matrix of the system to be identified. This novel SID approach is applied to the identification of a locally lifted system with the missing neighboring-state information. By fully exploiting the *two-layer* structure of the block Toeplitz matrices in the data equation of a lifted state-space system, a low-rank optimization problem is formulated for which the optimal solution can yield (parts of) the true Markov parameters of the locally lifted system. *Second* is the retrieval of

C. Yu and M. Verhaegen are with the Delft Center for Systems and Control, Delft University, Delft 2628CD, Netherlands (c.yu-4@tudelft.nl, m.verhaegen@tudelft.nl)

A. Hansson is with the Division of Automatic Control, Department of Electrical Engineering, Linkoping University, Sweden (anders.g.hansson@liu.se)

Part of the research of M. Verhaegen was done while he was a visiting professor of the Division of Automatic Control, Department of Electrical Engineering, Linkoping University, Sweden.

The research leading to these results has received funding from the European Research Council under the European Union's Seventh Framework Programme (FP7/2007-2013) / ERC grant agreement no. 339681.

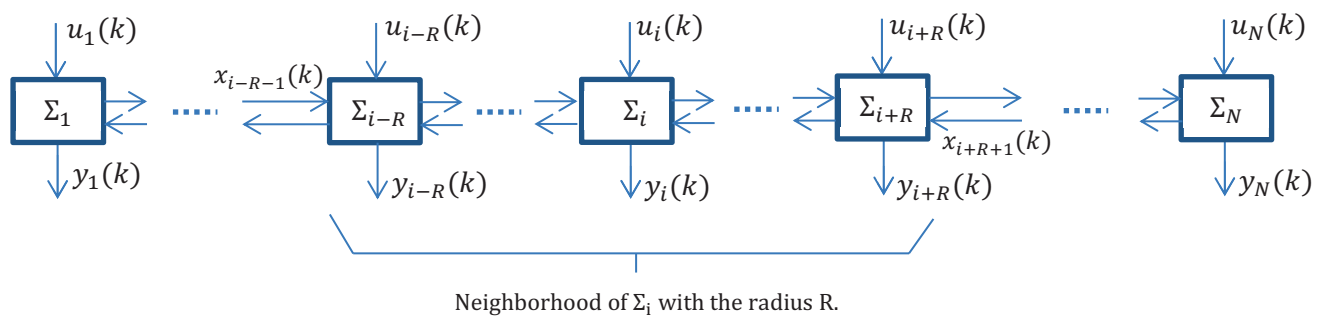

Fig. 1. Illustration of a cluster of subsystems in a neighborhood of the subsystem $\Sigma_i$ with radius $R$. The states $x_{i-R-1}(k)$ and $x_{i+R+1}(k)$ are explicitly indicated. They are like all other states unmeasurable.

the system matrices, describing the local LTI dynamics, of an individual subsystem *and* its interaction with its neighboring systems (up to a similarity transformation) from the reliably estimated Markov parameters. The retrieval of the system matrices of an individual subsystem is inherently a challenging structured state-space realization problem [18] for which the optimal solution can yield the estimates of system matrices up to a similarity transformation. The second feature also includes the exploitation of the shifting structure of a time-varying generalized observability matrix.

The rest of the paper is organized as follows. Section II describes the concerned identification problems and shows the challenge of dealing with the identification of a small cluster of subsystems in a large-scale networked system. Section III shows that the existing subspace identification methods break down in dealing with the concerned identification problem. Section IV presents a method for identifying the Markov parameters of the locally lifted state-space models. Section V provides a solution to the state-space realization of a single subsystem. Section VI provides a numerical example. The conclusions are provided in Section VII.

The following notations are adopted throughout the paper. The lowercase (uppercase) $x$ ($X$) is used to denote a vector (matrix). The superscripts $^T$ and $^{-1}$ are transpose and inverse operators, respectively. $\mathrm{vec}\,(X)$ represents the vector stacked by columns of $X$. $\mathrm{tr}\,(X)$ stands for the trace of $X$. $I_n$ denotes an $n \times n$ identity matrix and $O_{m,n}$ is an $m \times n$ zero matrix. $\otimes$ stands for the Kronecker product. $\|x\|_2$ represents the norm of $x$. $\|X\|_F$ and $\|X\|_*$ denote the Frobenius norm and nuclear norm of the matrix $X$, respectively.

## II. Preliminaries and Problem Definition

We consider the LTI systems $\{\Sigma_i\}_{i=1}^N$ connected in a homogeneous 1D network as shown in Fig. 1. The lifted state-space model of the concerned network is written as:

$$\begin{array}{llcl}
\Sigma_1: & x_1(k+1) & = & Ax_1(k) + A_r x_2(k) + Bu_1(k) \\
 & y_1(k) & = & Cx_1(k) + e_1(k) \\
\Sigma_i: & x_i(k+1) & = & Ax_i(k) + A_l x_{i-1}(k) + A_r x_{i+1}(k) \\
 & & & +Bu_i(k) \\
 & y_i(k) & = & Cx_i(k) + e_i(k) \\
\multicolumn{4}{l}{i = 2, \cdots, N-1,} \\
\Sigma_N: & x_N(k+1) & = & Ax_N(k) + A_l x_{N-1}(k) + Bu_N(k) \\
 & y_N(k) & = & Cx_N(k) + e_N(k)
\end{array} \tag{1}$$

where $x_i(k) \in \mathbb{R}^{n\times 1}$, $u_i(k) \in \mathbb{R}^{m\times 1}$, $y_i(k) \in \mathbb{R}^{p\times 1}$ and $e_i(k) \in \mathbb{R}^{p\times 1}$ are the state, input, output and measurement noise of the $i$-th subsystem, $x_{i-1}(k)$ and $x_{i+1}(k)$ are the neighboring states of the $i$-th subsystem. For $x_i(k)$, the subscript $i$ denotes the spatial index and $k$ is referred to as the time index. For a large-scale distributed system, we always assume that $N \gg n$ and $n > \max\{p, m\}$.

By lifting all states $x_i(k)$ into the vector $x(k)$ as $x(k) = \begin{bmatrix} x_1^T(k) & \cdots & x_N^T(k) \end{bmatrix}^T$ and doing the same for the inputs, outputs and noises defining resp. the vectors $u(k)$, $y(k)$ and $e(k)$, the global system in Fig. 1 has the following state space model:

$$\begin{aligned} x(k+1) &= \mathcal{A}x(k) + \mathcal{B}u(k) \\ y(k) &= \mathcal{C}x(k) + e(k), \end{aligned} \tag{2}$$

where $\mathcal{A}, \mathcal{B}, \mathcal{C}$ are $N \times N$ block matrices which have the following forms:

$$\mathcal{A} = \begin{bmatrix} A & A_r & & \\ A_l & A & \ddots & \\ & \ddots & \ddots & A_r \\ & & A_l & A \end{bmatrix},$$

$$\mathcal{B} = \begin{bmatrix} B & & & \\ & B & & \\ & & \ddots & \\ & & & B \end{bmatrix}, \ \mathcal{C} = \begin{bmatrix} C & & & \\ & C & & \\ & & \ddots & \\ & & & C \end{bmatrix}.$$

To identify the system model in (2), the existing SID methods only estimate the triplet $(\mathcal{A}, \mathcal{B}, \mathcal{C})$ *up to a similarity transformation*, thereby not preserving the special block-diagonal and block-tridiagonal structure of these system matrices. In addition, the computational complexity of SID methods, which is at least $O(N^3)$ with $N$ being the number of subsystems in the network, may easily disqualify their use for the identification of large-scale networks.

To deal with the high computational complexity for identifying the global system model in (2), we consider the identification of a cluster of subsystems $\{\Sigma_j\}_{j=i-R}^{i+R}$ in a neighborhood of $\Sigma_i$ with radius $R$ satisfying $R < i < N - R$ and $R \ll N$, as shown in Fig. 1. The lifted state-space model of this cluster is represented as

$$\begin{aligned} \underline{x}_i(k+1) &= \underline{A}_R \underline{x}_i(k) + \underline{B}_R \underline{u}_i(k) + \underline{D}_R \underline{v}_i(k), \\ \underline{y}_i(k) &= \underline{C}_R \underline{x}_i(k) + \underline{e}_i(k), \end{aligned} \tag{3}$$

with $\underline{x}_i(k), \underline{u}_i(k), \underline{y}_i(k)$ the subparts of $x(k), u(k), y(k)$ respectively from the block-rows $(i-R)$ to $(i+R)$; $\underline{A}_R, \underline{B}_R, \underline{C}_R$ are $(2R+1) \times (2R+1)$ block matrices which have forms similar to $\mathcal{A}, \mathcal{B}, \mathcal{C}$ in (2), respectively; the $(2R+1) \times 2$ block matrix $\underline{D}_R$ and the vector $\underline{v}_i(k)$ are defined as:

$$\underline{D}_R = \begin{bmatrix} A_l & 0 \\ 0 & 0 \\ \vdots & \vdots \\ 0 & 0 \\ 0 & A_r \end{bmatrix}, \quad \underline{v}_i(k) = \begin{bmatrix} x_{i-R-1}(k) \\ x_{i+R+1}(k) \end{bmatrix}.$$

For the identification of the cluster model in (3), the following two problems will be addressed.

**Problem 1:** The Markov parameters of the locally lifted state-space model in (3) will be identified using only the local input $\underline{u}_i(k)$ and the local output $\underline{y}_i(k)$.

This problem will be addressed in Section IV. The difficulty of this problem w.r.t. classical subspace identification methods in [19], [20] or their more recent variants [21], [22] is that the adjacent state sequences $\underline{x}_{i-R-1}(k)$ and $\underline{x}_{i+R+1}(k)$ are missing or the input sequence $\underline{v}_i(k)$ of the considered cluster model is unknown.

**Problem 2:** The order the local subsystem $\Sigma_i$, denoted by $n$, *as well as* the local system matrices $A, A_r, A_l, B, C$ are to be identified based on the identified Markov parameters in Problem 1. More specifically, the system matrices are to be identified up to a similarity transformation, i.e., the estimates $\hat{A}, \hat{A}_r, \hat{A}_l, \hat{B}, \hat{C}$ of $A, A_r, A_l, B, C$ satisfy $\hat{A} = Q^{-1}AQ,\ \hat{A}_l = Q^{-1}A_lQ,\ \hat{A}_r = Q^{-1}A_rQ,\ \hat{B} = Q^{-1}B,\ \hat{C} = CQ$ with $Q \in \mathbb{R}^{n\times n}$ being a non-singular ambiguity matrix.

The second problem will be addressed in Section V. It is noteworthy that, when the system order $n$ is known, the second problem is inherently a structured state-space realization problem as studied in [18], which turns out to be a challenging non-convex optimization problem.

The developed method in this paper can be applied to more general classes of large-scale identification problems than just a 1D chain of identical LTI systems, such as 1D heterogeneous networks consisting of clusters of identical LTI dynamical systems [23].

In addressing the locally lifted identification problem in (3), we stipulate the following assumption.

**Assumption A.1** The global system $(\mathcal{A}, \mathcal{B}, \mathcal{C})$ and the locally lifted system $(\underline{A}_R, \underline{B}_R, \underline{C}_R)$ are assumed to be minimal.

The persistent excitation of the input signal $u(k)$, which will be used for the identifiability analysis in the sequel, is defined below.

**Definition 1.** *A time sequence $u(k) \in \mathbb{R}^{Nm}$ is persistently exciting of order $s$ if there exists an integer $h$ such that the (block-) Hankel matrix*

$$\begin{bmatrix} u(k) & u(k+1) & \cdots & u(k+h-1) \\ u(k+1) & u(k+2) & \cdots & u(k+h) \\ \vdots & \vdots & \iddots & \vdots \\ u(k+s-1) & u(k+s) & \cdots & u(k+s+h-2) \end{bmatrix},$$

*has full row rank for any positive integer $k$.*

## III. Break down of existing subspace identification methods

The data equation of the local state-space model (3) is given as follows:

$$Y^i_{s,h} = \mathcal{O}_s x^i_h + \mathcal{T}^{B_R}_s U^i_{s,h} + \mathcal{T}^{D_R}_s V^i_{s,h} + E^i_{s,h}, \tag{4}$$

In this equation, the (block-) Hankel matrix $Y^i_{s,h}$ is defined as

$$Y^i_{s,h} = \begin{bmatrix} \underline{y}_i(1) & \cdots & \underline{y}_i(h) \\ \vdots & \iddots & \vdots \\ \underline{y}_i(s) & \cdots & \underline{y}_i(h+s-1) \end{bmatrix}$$

with the superscript $i$ being the spatial index of the subsystem $\Sigma_i$, the subscripts $s, h$ respectively being the number of block rows and the number of block columns. Analogous to $Y^i_{s,h}$, we define the block-Hankel matrices $U^i_{s,h}$, $V^i_{s,h}$, $E^i_{s,h}$ from the sequences $\underline{u}_i(k), \underline{v}_i(k), \underline{e}_i(k)$, respectively. The matrix $\mathcal{T}^{B_R}_s$ is a block Toeplitz matrix defined from the triplet $(\underline{A}_R, \underline{B}_R, \underline{C}_R)$ as

$$\mathcal{T}^{B_R}_s = \begin{bmatrix} 0 & & & \\ \underline{C}_R\underline{B}_R & 0 & & \\ \vdots & \ddots & \ddots & \\ \underline{C}_R\underline{A}_R^{s-2}\underline{B}_R & \cdots & \underline{C}_R\underline{B}_R & 0 \end{bmatrix},$$

and $\mathcal{T}^{D_R}_s$ is defined in a similar way from the triplet $(\underline{A}_R, \underline{D}_R, \underline{C}_R)$. The final matrix definitions in (4) are

$$\mathcal{O}_s = \begin{bmatrix} \underline{C}_R \\ \underline{C}_R\underline{A}_R \\ \vdots \\ \underline{C}_R\underline{A}_R^{s-1} \end{bmatrix}$$

and

$$x^i_h = [\underline{x}_i(k) \cdots \underline{x}_i(k+h-1)]\,.$$

*Existing subspace identification (SID) methods, see e.g. [17], break down when retrieving a matrix with a subspace like $\mathcal{O}_s$ or $x^i_h$ from the available data matrices $Y^i_{s,h}$ and $U^i_{s,h}$ in* (4).

To explain this deficiency of existing SID methods, we consider the simple noise-free case $e(k) \equiv 0$. Let

$$\Pi^\perp_U = I - U^{i,T}_{s,h}\left(U^i_{s,h}U^{i,T}_{s,h}\right)^{-1}U^i_{s,h},$$

Then the subspace revealing matrix [17] is $Y^i_{s,h}\Pi^\perp_U$, or $\begin{bmatrix} Y^{i,T}_{s,h} & U^{i,T}_{s,h} \end{bmatrix}^T$ if we consider the approach in [22].

When the unknown system input $\underline{v}_i(k)$ in (3) is absent, namely $\underline{v}_i(k) \equiv 0$ (or $V^i_{h,s} \equiv 0$), the column space of $\mathcal{O}_s$ can be retrieved from that of $Y^i_{s,h}\Pi^\perp_U$, while the row space of $x^i_h$ can be retrieved from that of compound matrix $\begin{bmatrix} Y^{i,T}_{s,h} & U^{i,T}_{s,h} \end{bmatrix}^T$. However, for the case $\underline{v}_i(k) \neq 0$, the presence of the bilinear term $\mathcal{T}^{D_R}_s V^i_{s,h}$ destroys the important subspace revealing property in existing SID methods.

Next, we shall show that the identification method in [21] cannot handle the concerned identification problem. One could consider the unknown input $\underline{v}_i(k)$ in (3) as a "missing input variable". However, for the simple case $e(k) \equiv 0$ and using the notation $\mathcal{H}_v$ to denote the set block Hankel matrices of the same structure as the matrix $V^i_{s,h}$, the rank minimization problem of [21] can be written as

$$\min_{V\in\mathcal{H}_v} \quad \text{rank}\begin{bmatrix} Y^i_{s,h} \\ U^i_{s,h} \\ V \end{bmatrix}. \tag{5}$$

By the following inequality

$$\text{rank}\begin{bmatrix} Y^i_{s,h} \\ U^i_{s,h} \\ V \end{bmatrix} \geq \text{rank}\begin{bmatrix} Y^i_{s,h} \\ U^i_{s,h} \end{bmatrix},$$

we can see that $V = 0$ is an optimal solution to (5). It implies that the whole missing input sequence, not just a few missing entries, cannot be recovered using the low-rank optimization of [21].

To resolve this deficiency of SID methods, when considering the identification of local clusters in a large homogeneous network, a novel SID approach is developed in this paper. This method differs from the existing SID methods in two major ways.

First, the low-rank subspace retrieval step that is characteristic for many SID methods is unable to find an accurate (or consistent) estimate of the key subspace of $\mathcal{O}_s$ or $x_h^i$. In this paper, the low-rank optimization will be used for the accurate estimation of (parts of) the Markov parameters in the block Toeplitz matrix $\mathcal{T}_s^{B_R}$.

Second, contrary to existing SID methods that are unable to preserve the structures in the estimated system matrices, the proposed subspace identification method is capable of accurately estimating the non-zero block entries (up to a similarity transformation) of the structured system matrices in (3), as stated in Problem 2.

## IV. Identifying the Markov parameters of the locally lifted state space model (3)

In this section, the Markov parameters of the state-space model in (3) will be estimated by exploiting the low-rank property of the unmeasurable-state related terms in (4) and the specific block Toeplitz structure of $\mathcal{T}_s^{B_R}$. Subsection IV-A formulates a low rank optimization problem without considering the specific structure of $\mathcal{T}_s^{B_R}$. It is shown in Lemma 2 that this low-rank optimization is unable to recover the true Markov parameters. Subsection IV-B reformulates another low-rank optimization problem by incorporating the specific Toeplitz structure of $\mathcal{T}_s^{B_R}$. It is shown in Theorem 1 that, under some mild conditions, the optimal solution to the proposed low-rank optimization problem can yield (parts of) the true Markov parameters.

### A. Formulation of a low-rank optimization problem

Due to the unmeasurable state sequences in a network, the matrix sum $\mathcal{O}_s x_h^i + \mathcal{T}_s^{D_R} V_{s,h}^i$ in (4) is unknown. However, this matrix sum has a low-rank property that will be exploited as a solution in the new subspace identification method.

**Lemma 1.** *For the data equation* (4)*, when $h > ps$ or $Y_{s,h}^i$ is a fat matrix, the sum $\mathcal{O}_s x_h^i + \mathcal{T}_s^{D_R} V_{s,h}^i$ satisfies the following rank property*

$$\text{rank}\left(\mathcal{O}_s x_h^i + \mathcal{T}_s^{D_R} V_{s,h}^i\right) \leq (2R+1)n + \min\{(s-1)sp, 2(s-1)n\}, \tag{6}$$

*where $(s-1)sp$ and $2(s-1)n$ denote the number of non-zero rows and columns of $\mathcal{T}_s^{D_R}$, respectively.*

*Proof:* From the structures of $\mathcal{O}_s$ and $\mathcal{T}_s^{D_R}$, we can get that $\text{rank}\left(\mathcal{O}_s x_h^i\right) \leq \text{rank}\left(\mathcal{O}_s\right) \leq (2R+1)n$ and $\text{rank}\left(\mathcal{T}_s^{D_R} V_{s,h}^i\right) \leq \text{rank}\left(\mathcal{T}_s^{D_R}\right) \leq \min\{(s-1)sp, 2(s-1)n\}$. Thus, the result in the lemma is straightforward. ■

From Lemma 1, we can derive a condition to select the parameter $s$ in the data equation (4) and the cluster radius $R$, defined above equation (3), such that the sum $\mathcal{O}_s x_h^i + \mathcal{T}_s^{D_R} V_{s,h}^i$ is of low rank (or rank deficient). This condition reads:

$$(2R+1)sp > (2R+1)n + \min\{(s-1)sp, 2(s-1)n\}. \tag{7}$$

The above condition means that the number of the rows of the matrix $\mathcal{O}_s x_h^i + \mathcal{T}_s^{D_R} V_{s,h}^i$ is larger than an upper bound of its rank. In practice, by fixing a value of $s$ satisfying that $s > \frac{n}{p}$, we can always find a value of $R$ such that the above inequality holds. Therefore, in the sequel, we assume that the matrix sum $\mathcal{O}_s x_h^i + \mathcal{T}_s^{D_R} V_{s,h}^i$ is of low-rank (or rank deficient).

Denote the noise-free output as $\hat{y}_i(k) = \underline{y}_i(k) - \underline{e}_i(k)$ and its related block Hankel matrix $\hat{Y}_{s,h}^i$. Based on the rank property discussed above, a low-rank regularized optimization problem is then proposed as follows:

$$\begin{aligned} \min_{\Theta_s^{B_R} \in \mathcal{T}, \hat{Y}_{s,h}^i \in \mathcal{H}} & \sum_{t=1}^{h+s-1} \|\hat{y}_i(t) - \underline{y}_i(t)\|^2 \\ & + \lambda \cdot \text{rank}\left[\hat{Y}_{s,h}^i - \Theta_s^{B_R} U_{s,h}^i\right], \end{aligned} \tag{8}$$

where $\mathcal{T}$ and $\mathcal{H}$ denote respectively the set of block Toeplitz and block Hankel matrices with appropriate block sizes, and the regularization parameter $\lambda$ allows to make a trade-off between the two terms in the cost function. It is remarked that the block Toeplitz structure of $\mathcal{T}$ corresponds to the first block-Toeplitz layer that is described in Subsection IV-B.

In the absence of measurement noise, it will be shown in the following lemma that the optimal solution to (8) is non-unique.

**Lemma 2.** *Consider the optimization problem in* (8)*. Suppose that the following assumptions are satisfied:*

1) *The global system input $u(k)$ to* (2) *is persistently exciting of order $Nn+s$ with $s$ being the SID dimension parameter in* (4)*;*
2) *The measurement noise is absent, i.e., $\hat{y}_i(k) = \underline{y}_i(k)$;*
3) *The matrix pairs $(A_l, B)$ and $(A_r, B)$ have full row rank.*

*Then, the optimal solution to the following rank optimization problem is non-unique:*

$$\min_{\Theta_s^{B_R} \in \mathcal{T}} \text{rank}\left[Y_{s,h}^i - \Theta_s^{B_R} U_{s,h}^i\right] \quad \text{for } R+s < i < N-R-s. \tag{9}$$

The proof to the above lemma is provided in Appendix A. The above lemma indicates that the Markov parameters, as the block entries of $\mathcal{T}_s^{B_R}$, cannot be recovered even if the global optimal solution to the low-rank optimization problem in (9) can be found.

In the next subsection, a solution will be provided by further constraining the structure of block entries of $\mathcal{T}_s^{B_R}$, due to the specific structures of the system matrices $(\underline{A}_R, \underline{B}_R, \underline{C}_R)$ in (3).

### B. Structure constrained low-rank optimization

The block matrix $\mathcal{T}_s^{B_R}$ in (4) has a two-layer block Toeplitz structure, which will be utilized in the proposed identification method. The first layer is the block Toeplitz structure of $\mathcal{T}_s^{B_R}$ with respect to its block entries $\underline{C}_R\underline{A}_R^j\underline{B}_R$. The second layer is the partial block Toeplitz structure inside the block entries $\underline{C}_R\underline{A}_R^j\underline{B}_R$, as highlighted in the following example and lemma.

**Example 1.** *If we take $R = 3$ and assume that each block in $\underline{A}_R, \underline{B}_R, \underline{C}_R$ has size $2 \times 2$, then a visual illustration of the structures of the matrices $\{\underline{M}_j = \underline{C}_R\underline{A}_R^j\underline{B}_R\}_{j=1}^3$ is given in Figure 2.*

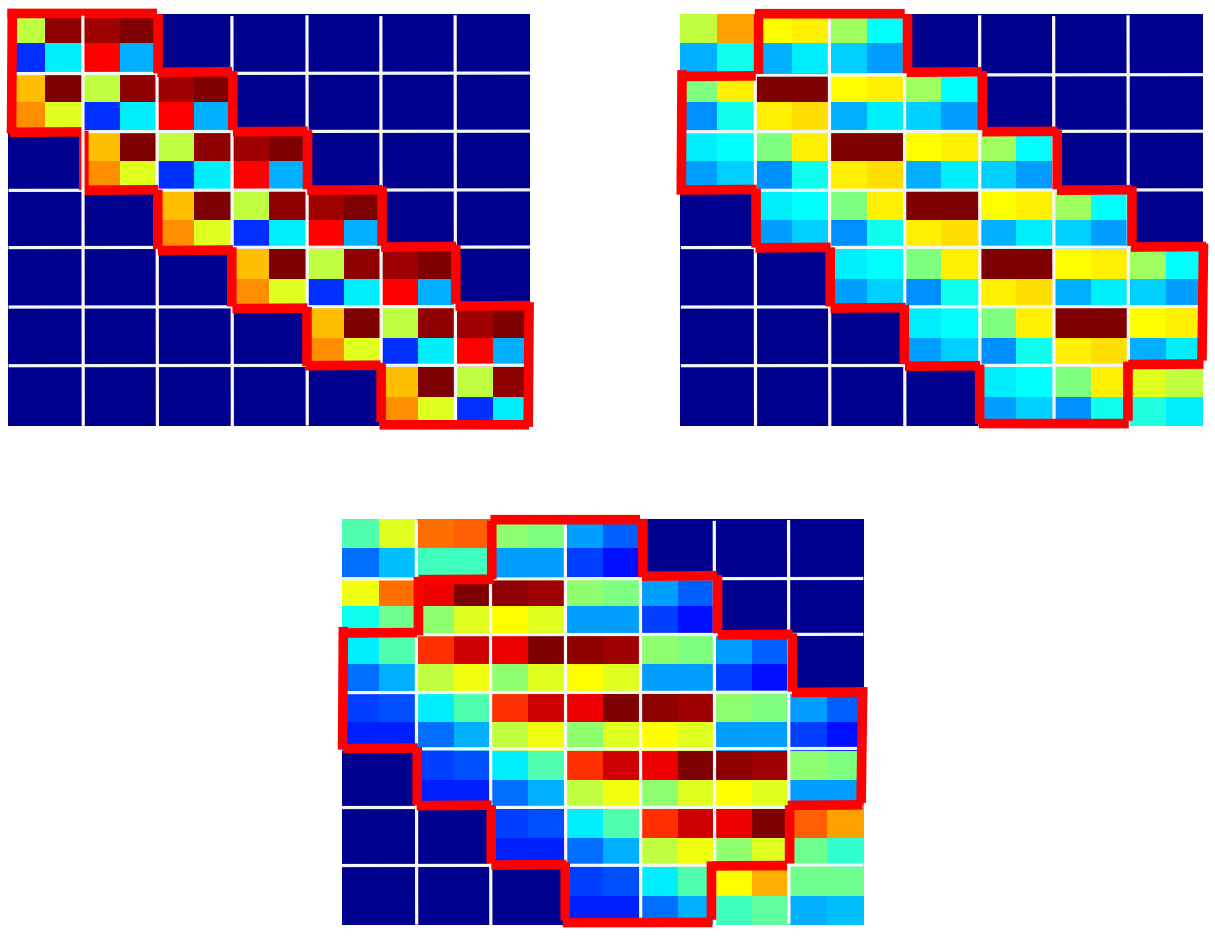

Fig. 2. Illustration of the partial Toeplitz structure of the matrices $\{\underline{M}_j = \underline{C}_R\underline{A}_R^j\underline{B}_R\}_{j=1}^3$ Top-left for $j = 1$ with block-bandwidth 1; top-right for $j = 2$ with block-bandwidth 2; bottom for $j = 3$ with block-bandwidth 3. The deep blue color represents zero entries. The parts surrounded by red curves have block Toeplitz structures.

**Lemma 3.** *Based on the matrices $\underline{A}_R, \underline{B}_R, \underline{C}_R$ defined in* (3)*, the following hold about the matrix product $\underline{M}_j = \underline{C}_R\underline{A}_R^j\underline{B}_R$:*

1) *$\underline{M}_j$ is a banded block matrix, with block-bandwidth $j$.*
2) *The submatrices of $\underline{M}_j$ , for $j < 2R + 1$, at the $l-$th block row and $q-$th block column with $l, q \in \{1, \cdots, 2R + 1\}$, are inside the partial block-Toeplitz region for the index-pair $(l, q)$ satisfying:*

$$i + 1 \le l + q \le 4R + 3 - i.$$

*Proof:* Since $\underline{B}_R$ and $\underline{C}_R$ are block diagonal matrices with constant diagonal blocks, $\underline{M}_j$ has the same structure pattern as $\underline{A}_R^j$. The matrix $\underline{A}_R$ can be written as

$$\underline{A}_R = I \otimes A + J_- \otimes A_l + J_+ \otimes A_r$$

with $J_-$ and $J_+$ block-column shifting matrices such that post-multiplication of a matrix with $J_-$ results in shifting the block-columns of that matrix to the left and adding a zero block column in the last column, while post-multiplication of a matrix with $J_+$ results in the opposite shift and adding a zero block column in the first column.

The proof is completed by induction. For $j = 1$ and $j = 2$ this can be easily checked. Let us assume that $\underline{A}_R^j$ has a block Toeplitz structure for the matrix blocks with the index-pair $(l, q)$ satisfying that $j + 1 \le l + q \le 4R + 3 - j$, and using the above expression for $\underline{A}_R$, we can express $\underline{A}_R^{j+1}$ as

$$\underline{A}_R^{j+1} = \underline{A}_R^j\underline{A}_R = \underline{A}_R^j(I \otimes A + J_- \otimes A_l + J_+ \otimes A_r). \quad (10)$$

By several trivial manipulations of the above matrix product, it can be derived that $\underline{A}_R^{j+1}$ has a partial Toeplitz structure for the matrix blocks with the index-pair $(l, q)$ satisfying that $(j + 1) + 1 \le l + q \le 4R + 3 - (j + 1)$. This completes the proof. ■

Let $\mathcal{T}_f$ denote the set of two-layer block Toeplitz matrices that was outlined above for the matrix $\mathcal{T}_s^{B_R}$. More clearly, the structure of $\mathcal{T}_s^{B_R}$ is illustrated in the following example.

**Example 2.** *If we take $R = 2$ and $s = 4$, the Toeplitz matrix $\mathcal{T}_s^{B_R}$ can be parameterized in terms of the parameter set $\{\varphi_j\}_{j=1}^{11}$ with $\varphi_j \in \mathbb{R}^{p\times m}$ as follows:*

$$\mathcal{T}_s^{B_R} = \begin{bmatrix} 0 & & & \\ \underline{M}_0 & 0 & & \\ \underline{M}_1 & \underline{M}_0 & 0 & \\ \underline{M}_2 & \underline{M}_1 & \underline{M}_0 & 0 \end{bmatrix}, \quad (11)$$

*where* $$\underline{M}_0 = \begin{bmatrix} \varphi_1 & & & & \\ & \varphi_1 & & & \\ & & \varphi_1 & & \\ & & & \varphi_1 & \\ & & & & \varphi_1 \end{bmatrix},$$

$$\underline{M}_1 = \begin{bmatrix} \varphi_3 & \varphi_4 & & & \\ \varphi_2 & \varphi_3 & \varphi_4 & & \\ & \varphi_2 & \varphi_3 & \varphi_4 & \\ & & \varphi_2 & \varphi_3 & \varphi_4 \\ & & & \varphi_2 & \varphi_3 \end{bmatrix}, \quad \underline{M}_2 = \begin{bmatrix} \varphi_{10} & \varphi_8 & \varphi_9 & & \\ \varphi_6 & \varphi_7 & \varphi_8 & \varphi_9 & \\ \varphi_5 & \varphi_6 & \varphi_7 & \varphi_8 & \varphi_9 \\ & \varphi_5 & \varphi_6 & \varphi_7 & \varphi_8 \\ & & \varphi_5 & \varphi_6 & \varphi_{11} \end{bmatrix}.$$

Then we specialize the rank-optimization problem in (8) as

$$\min_{\Theta_s^{B_R}\in\mathcal{T}_f, \hat{Y}_{s,h}^i\in\mathcal{H}} \sum_{t=1}^{h+s-1} \|\hat{y}_i(t) - \underline{y}_i(t)\|^2 + \lambda \cdot \mathrm{rank}\left[\hat{Y}_{s,h}^i - \Theta_s^{B_R}U_{s,h}^i\right]. \quad (12)$$

The difference between (8) and (12) is in the definition of the constrained Toeplitz sets indicated by $\mathcal{T}$ and $\mathcal{T}_f$, respectively. Though it is only a minor notational difference, the impact on the solution is huge. As will be outlined in Theorem 1, it enables uniqueness of part of the solution.

For the uniqueness property, use will be made of the following time-varying observability matrix $\mathbf{O}_{j,k}$, which is a sub-matrix of the extended observability matrix $\mathcal{O}_j$ defined in (4), consisting of the block rows corresponding to the second-layer block Toeplitz part of $\mathcal{T}_s^{B_R}$.

**Definition 2.** *Let $G_j$ be a $j \times (j + 2)$ block Toeplitz matrix of the form*

$$G_j = \begin{bmatrix} A_l & A & A_r & & & \\ & A_l & A & A_r & & \\ & & \ddots & \ddots & \ddots & \\ & & & A_l & A & A_r \end{bmatrix}. \quad (13)$$

*Denote $\underline{C}_j = I_j \otimes C$. A time-varying observability matrix $\mathbf{O}_{j,k}$, for $j > 2(k-1)$, is defined in terms of the matrix pair $(\underline{C}_j, G_j)$ as [24, Chapter 3]:*

$$\mathbf{O}_{j,k} = \begin{bmatrix} \underline{C}_j \\ \underline{C}_{j-2}G_{j-2} \\ \underline{C}_{j-4}G_{j-4}G_{j-2} \\ \vdots \\ \underline{C}_{j-2(k-1)}G_{j-2(k-1)}\cdots G_{j-2} \end{bmatrix}.$$

**Theorem 1.** *Suppose that the following assumptions are satisfied:*

1) *Assumption A.1 holds and $\nu_o$ is the observability index of the pair $(\underline{A}_R, \underline{C}_R)$;*
2) *The cluster* radius *$R$ of the lifted model in* (3) *and the dimension parameter $s$ in* (4) *satisfy*
$$s > \nu_o, \quad R \geq s-1;$$
3) *The conditions 1)-3) of Lemma 2 hold;*
4) *The time-varying observability matrix $\mathbf{O}_{2R+1,s-1}$, defined in **Definition** 2, has full column rank.*

*Then the submatrices of the Markov parameters $\underline{M}_j = \underline{C}_R\underline{A}_R^j\underline{B}_R$, for $j = 0, 1, \cdots, s-2$, contained in the matrix $\mathcal{T}_s^{B_R}$ in* (4)*, for which Lemma 3 has shown that they preserve the second layer of block-Toeplitz structure, can be computed in a unique manner from the following low-rank optimization problem:*

$$\min_{\Theta_s^{B_R} \in \mathcal{T}_f} \text{rank}\left[Y_{s,h}^i - \Theta_s^{B_R}U_{s,h}^i\right] \quad \text{for } R+s < i < N-R-s. \tag{14}$$

The proof of the above theorem can be found in the Appendix B, where we can see that the (unique) global optimal solution can yield correct recovery of certain parts of the Markov parameters $\underline{M}_j = \underline{C}_R\underline{A}_R^j\underline{B}_R$, for $j = 0, 1, \cdots, s-2$. In addition, based on the conditions 2) and 4) of the above theorem, it can be established that the inequality (7) holds, namely the matrix sum $\mathcal{O}_s x_h^i + \mathcal{T}_s^{D_R}V_{s,h}^i$ is indeed of low-rank (or rank deficient). This in turn indicates that the low-rank cost function in (12) is a reasonable choice.

Lemma 2 and Theorem 1 demonstrate that the second-layer Toeplitz structure, as pointed out in Lemma 3, is crucial for enforcing a (unique) solution to the identification problem 1 of this paper. The considered identification of the Markov parameters boils down to solving the low-rank regularized optimization problem in (12).

Since the optimization problem in (12) is non-convex, it is difficult to obtain an optimal solution under a mild computational burden. In this paper, the reweighted nuclear norm optimization method [25] is adopted, which is regarded as an iterative heuristic for the rank minimization problem (12).

## V. State-space realization of a single subsystem

In this section, we shall study the final realization of the system matrices $\{C, A, A_l, A_r, B\}$ from the estimated submatrices of the Markov parameters $\underline{M}_j = \underline{C}_R\underline{A}_R^j\underline{B}_R$, for $j = 0, 1, \cdots, s-2$, contained in the matrix $\mathcal{T}_s^{B_R}$ in (4), for which Lemma 3 has shown that they preserve the second-layer of block-Toeplitz structure.

**Remark 1.** *The realization problem considered in this section is a structured state-space realization problem, which turns out to be a challenging non-convex optimization problem [18], [20], [26]. This problem is usually solved by the gradient-based optimization methods, which are sensitive to the selection of the initial condition. Here, we transform the realization problem into a structured low-rank matrix factorization problem for which the optimal solution enables the estimation of system matrices $(C, A_l, A, A_r, B)$ up to a similarity transformation. This factorization problem is then reformulated into an equivalent low-rank optimization problem, which is then numerically solved using the optimization method developed in [27].*

We start the solution by developing expressions of the submatrices in the second-layer Toeplitz regions in terms of the system matrices $\{C, A, A_l, A_r, B\}$. This is done in the following Lemma.

**Lemma 4.** *Consider the block matrices $\underline{A}_R$, $\underline{B}_R$ and $\underline{C}_R$ defined in (3). Let the sequence of non-zero block entries from left to right of the $(j+1)$-th block row of the matrix $\underline{C}_R\underline{A}_R^j\underline{B}_R$ be denoted as $\{F_{j,-j}, F_{j,1-j}, \cdots, F_{j,j-1}, F_{j,j}\}$, then these matrix entries satisfy the following relationship:*

$$\sum_{k=-j}^{j} F_{j,k}z^{-k} = C(A_l z^{-1} + A + A_r z)^j B, \tag{15}$$

*where $z \in \mathbb{C}$.*

*Proof:* The above result can be derived using the filter bank theory in [28]. ■

As $F_{j,k}$ are the Markov parameters inside the second-layer block Toeplitz part of $\mathcal{T}_s^{B_R}$, the values of $F_{j,k}$ for $j \in \{0, 1, \cdots, s-2\}, k \in \{-j, \cdots, j\}$ are assumed to be available in this section. Based on these values $F_{j,k}$, we address the problem of estimating the matrices $\{C, A, A_l, A_r, B\}$ up to a similarity transformation.

Dual to Definition 2, we shall define a time-varying controllability matrix $\mathbf{C}_{j,k}$, which is a sub-matrix of the extended controllability matrix determined by $(\underline{A}_R, \underline{B}_R)$.

**Definition 3.** *Let $\Gamma_j$ be a $(j+2) \times j$ block Toeplitz matrix of the form*

$$\Gamma_j = \begin{bmatrix} A_r & & & \\ A & A_r & & \\ A_l & A & \ddots & \\ & A_l & \ddots & A_r \\ & & \ddots & A \\ & & & A_l \end{bmatrix}. \tag{16}$$

*Denote $\underline{B}_j = I_j \otimes B$. A time-varying controllability matrix $\mathbf{C}_{j,k}$, for $j > 2(k-1)$, is defined in terms of the matrix pair $(\Gamma_j, \underline{B}_j)$:*

$$\mathbf{C}_{j,k} = \left[\begin{array}{c|c|c|c} \underline{B}_j & \Gamma_{j-2}\underline{B}_{j-2} & \cdots & \Gamma_{j-2}\cdots\Gamma_{j-2(k-1)}\underline{B}_{j-2(k-1)} \end{array}\right].$$

In the sequel, we let $s$ be an even integer such that $s/2$ is an integer as well. The solution to the realization of the system matrices $\{C, A_l, A, A_r, B\}$ is done in two phases. In the first phase, the structured time-varying observability matrix $\mathbf{O}_{2R+1,s/2}$ and the structured time-varying controllability matrix $\mathbf{C}_{2R+1,s/2}$ are to be estimated from the available matrix values $F_{i,j}$. In the second phase, the system matrices $\{C, A, A_l, A_r, B\}$ are derived from these time-varying observability and controllability matrices. It is remarked that the subscript $_{s/2}$ in $\mathbf{O}_{2R+1,s/2}$ (or $\mathbf{C}_{2R+1,s/2}$) means that the maximum moment of the block entries in $\mathbf{O}_{2R+1,s/2}$ (or $\mathbf{C}_{2R+1,s/2}$) is $s/2-1$. The subscript $_{s/2}$ is adopted because the sum of the maximum moments of $\mathbf{O}_{2R+1,s/2}$ and $\mathbf{C}_{2R+1,s/2}$ is equal to the maximum moment, $s-2$, of the available Markov parameters $\underline{M}_j$.

### *A. Determining the time-varying observability and controllability matrices*

In this subsection, the determination of $\mathbf{O}_{2R+1,s/2}$ and $\mathbf{C}_{2R+1,s/2}$ will be formulated as a structured low-rank matrix-factorization problem. More importantly, we show that the optimal solution to this matrix-factorization problem can yield the estimates of $\mathbf{O}_{2R+1,s/2}$ and $\mathbf{C}_{2R+1,s/2}$ up to a block-diagonal ambiguity matrix with identical block-diagonal entries. This is crucial to identify the system matrices $\{C, A, A_l, A_r, B\}$ up to a similarity transformation, as stated in Problem 2.

First, by the definitions of $\mathbf{O}_{2R+1,s/2}$ and $\mathbf{C}_{2R+1,s/2}$, we can find that the product of $\mathbf{O}_{2R+1,s/2}$ and $\mathbf{C}_{2R+1,s/2}$ is equal to a matrix constructed from $\{F_{j,k}\}_{k=-j}^{j}$ for $j = 0, 1, \cdots, s-2$. This is demonstrated in the following example.

**Example 3.** *When $R = 1$ and $s = 4$, the product of $\mathbf{O}_{3,2}$ and $\mathbf{C}_{3,2}$ can be expressed as*

$$\mathbf{O}_{3,2}\mathbf{C}_{3,2} = \left[\begin{array}{ccc} C & 0 & 0 \\ 0 & C & 0 \\ 0 & 0 & C \\ \hline CA_l & CA & CA_r \end{array}\right] \left[\begin{array}{ccc|c} B & 0 & 0 & A_rB \\ 0 & B & 0 & AB \\ 0 & 0 & B & A_lB \end{array}\right] = \left[\begin{array}{ccc|c} F_{0,0} & 0 & 0 & F_{1,1} \\ 0 & F_{0,0} & 0 & F_{1,0} \\ 0 & 0 & F_{0,0} & F_{1,-1} \\ \hline F_{1,-1} & F_{1,0} & F_{1,1} & F_{2,0} \end{array}\right].$$

*The above equation provides a simple example, showing that the product of $\mathbf{O}_{3,2}$ and $\mathbf{C}_{3,2}$ can be expressed in terms of $\{F_{j,k}\}_{k=-j}^{j}$ for $j = 0, 1, 2$.*

Here, the product of $\mathbf{O}_{2R+1,s/2}$ and $\mathbf{C}_{2R+1,s/2}$ is represented as:

$$\mathbf{O}_{2R+1,s/2}\mathbf{C}_{2R+1,s/2} = \mathbf{H}_{2R+1,2R+1}, \tag{17}$$

where $\mathbf{H}_{2R+1,2R+1}$ is a $(2R + 1) \times (2R + 1)$ block matrix that is assumed to be known.

Given the matrix $\mathbf{H}_{2R+1,2R+1}$, the problem of interest is to determine $\mathbf{O}_{2R+1,s/2}$ and $\mathbf{C}_{2R+1,s/2}$ from equation (17). According to Definitions 2 and 3, we can see that $\mathbf{O}_{2R+1,s/2}$ and $\mathbf{C}_{2R+1,s/2}$ are structured matrices. These structures are instrumental to analyzing the properties of the optimal solution to (17), as shown in Theorem 2. In order to show these structures, use will be made of the following matrix definition.

For $j = 0, 1, \cdots, s/2$ and $z \in \mathbb{C}$, the matrix sequences $\{W_{j,l}\}_{l=-j}^{j}$ and $\{E_{j,l}\}_{l=-j}^{j}$ are defined as

$$\begin{aligned} \sum_{l=-j}^{j} W_{j,l} z^{-l} &= C(A_l z^{-1} + A + A_r z)^j \\ \sum_{l=-j}^{j} E_{j,l} z^{-l} &= (A_l z^{-1} + A + A_r z)^j B. \end{aligned} \tag{18}$$

From the defined quantities $W_{j,l}$ and $E_{j,l}$ in (18), the matrices $\mathbf{W}_j$ and $\mathbf{E}_j$ are defined as

$$\mathbf{W}_j = \begin{bmatrix} W_{0,0} \\ W_{1,-1} \\ W_{1,0} \\ W_{1,1} \\ W_{2,-2} \\ \vdots \\ W_{j-1,j-1} \end{bmatrix} \quad \mathbf{E}_j = \begin{bmatrix} E_{0,0}^T \\ E_{1,-1}^T \\ E_{1,0}^T \\ E_{1,1}^T \\ E_{2,-2}^T \\ \vdots \\ E_{j-1,j-1}^T \end{bmatrix}^T. \tag{19}$$

It can be seen that $\mathbf{O}_{j,k}$ (or $\mathbf{C}_{j,k}$) is a block matrix constructed from the block components of $\mathbf{W}_k$ (or $\mathbf{E}_k$). This is illustrated in the next example.

**Example 4.** *The time-varying observability matrix $\mathbf{O}_{5,3}$ is represented in terms of $\mathbf{W}_3$ as follows*

$$\mathbf{O}_{5,3} = \left[\begin{array}{ccccc} W_{0,0} & & & & \\ & W_{0,0} & & & \\ & & W_{0,0} & & \\ & & & W_{0,0} & \\ & & & & W_{0,0} \\ \hline W_{1,-1} & W_{1,0} & W_{1,1} & & \\ & W_{1,-1} & W_{1,0} & W_{1,1} & \\ & & W_{1,-1} & W_{1,0} & W_{1,1} \\ \hline W_{2,-2} & W_{2,-1} & W_{2,0} & W_{2,1} & W_{2,2} \end{array}\right].$$

Denote by $\mathcal{O}_{2R+1,s/2}$ the set of block matrices having the same structure as $\mathbf{O}_{2R+1,s/2}$, as illustrated in Example 4 , and $\mathcal{C}_{2R+1,s/2}$ the set of block matrices having the same structure as $\mathbf{C}_{2R+1,s/2}$. It is remarked that the definitions of both the sets $\mathcal{O}_{2R+1,s/2}$ and $\mathcal{C}_{2R+1,s/2}$ require knowledge of the system order $n$. We then propose the following structured low-rank matrix factorization problem:

$$\begin{aligned} \min_{\mathbf{O},\mathbf{C}} \quad & \|\mathbf{H}_{2R+1,2R+1} - \mathbf{OC}\|_F^2 \\ s.t. \quad & \mathbf{O} \in \mathcal{O}_{2R+1,s/2}, \mathbf{C} \in \mathcal{C}_{2R+1,s/2}. \end{aligned} \tag{20}$$

It will be shown in the following theorem that the optimal solution to (20) can yield the estimates of $\mathbf{O}$ or $\mathbf{C}$ up to a block-diagonal ambiguity matrix with identical block-diagonal entries.

**Theorem 2.** *Consider the optimization problem in* (20)*. Suppose that the following assumptions are satisfied:*

1) *The values of $R$ and $s$ satisfy $R \geq s - 2$, and $s$ is a positive even integer;*
2) *The system order of a subsystem, $n$, is known;*

3) *The matrices* $\mathbf{O}_{j,s/2}$ *and* $\mathbf{C}_{j,s/2}$*, for any* $j \geq \min\{2s-3, 2R\}$*, have full column and row rank, respectively;*
4) *The matrix* $\mathbf{H}_{2R+1,2R+1}$ *satisfying equation* (17) *is known exactly.*

*Then, any optimal solution pair* $\{\hat{\mathbf{O}}, \hat{\mathbf{C}}\}$ *to the optimization* (20) *satisfies*

$$\begin{aligned}\hat{\mathbf{O}} &= \mathbf{O}_{2R+1,s/2}\mathbf{Q}\\ \hat{\mathbf{C}} &= \mathbf{Q}^{-1}\mathbf{C}_{2R+1,s/2},\end{aligned} \tag{21}$$

*where* $\mathbf{Q} = I_{2R+1} \otimes Q$ *with* $Q \in \mathbb{R}^{n\times n}$ *being a nonsingular ambiguity matrix.*

The proof of this theorem is given in Appendix C.

### *B. Rank-constrained form of the optimization problem* (20)

In this subsection, the structured low-rank matrix factorization problem in (20) is reformulated into a rank-constrained optimization problem.

From the matrix quantities defined in (18) and (19), we can see that $\mathbf{O}_{2R+1,s/2}$ is affine in terms of $\mathbf{W}_{s/2}$, while $\mathbf{C}_{2R+1,s/2}$ is affine in terms of $\mathbf{E}_{s/2}$. Then, it can be established that product $\mathbf{O}_{2R+1,s/2}\mathbf{C}_{2R+1,s/2}$ is affine in terms of block entries of $\mathbf{W}_{s/2}\mathbf{E}_{s/2}$ [29]. This affine operator $\mathcal{H}(\cdot)$ is defined by

$$\mathbf{O}_{2R+1,s/2}\mathbf{C}_{2R+1,s/2} = \mathcal{H}(\mathbf{W}_{s/2}\mathbf{E}_{s/2}). \tag{22}$$

In the sequel, the affine operator $\mathcal{H}(\cdot)$ is assumed to be known.

Instead of $\mathbf{O}_{2R+1,s/2}$ and $\mathbf{C}_{2R+1,s/2}$, we regard the product $\mathbf{X} = \mathbf{W}_{s/2}\mathbf{E}_{s/2}$ as the variable to be determined. Then, the affine function $\mathcal{H}(\mathbf{X})$ works on the block entries of $\mathbf{X}$. By taking into account the low-rank property of the matrix product $\mathbf{W}_{s/2}\mathbf{E}_{s/2}$, we propose the following rank-constrained optimization problem:

$$\begin{aligned}\min_{\mathbf{X}} \quad & \|\mathbf{H}_{2R+1,2R+1} - \mathcal{H}(\mathbf{X})\|_F^2\\ s.t. \quad & \text{rank}(\mathbf{X}) = n,\end{aligned} \tag{23}$$

where $n$ is the system order of a subsystem in (1) and $\mathbf{X}$ has the same size as $\mathbf{W}_{s/2}\mathbf{E}_{s/2}$.

Next, we will show that the rank-constrained optimization problem in (23) is an equivalent formulation of (20). To show this result, the following lemma is required.

**Lemma 5.** *Suppose that* $\mathbf{O}_{j,s/2}$ *and* $\mathbf{C}_{j,s/2}$*, for* $j > s-2$*, have full column and row rank, respectively. Then,* $\mathbf{W}_{s/2}$ *and* $\mathbf{E}_{s/2}$ *have full column and row rank, respectively, and both ranks are* $n$*.*

*Proof:* Since $\mathbf{O}_{j,s/2}$ has full column rank, the block vector $\left[W_{0,0}^T, W_{1,-1}^T, \cdots, W_{s/2,-s/2}^T\right]^T$, which is constructed by stacking the non-zero block entries in the first block column of $\mathbf{O}_{j,s/2}$, has full column rank as well. Since this block vector is a sub-part of $\mathbf{W}_{s/2}$, it can be derived that $\mathbf{W}_{s/2}$ has full column rank. Similarly, it can be proven that the block vector $\mathbf{E}_{s/2}$ has full row rank as well.

■

The next theorem shows the equivalence between the optimization problems in (20) and (23).

**Theorem 3.** *Consider the optimization problems in* (20) *and* (23)*. Suppose that all the conditions of Theorem 2 hold. Then, the optimization problems* (20) *and* (23) *are equivalent in the sense that their optimal solutions can yield the estimate* $\left(\hat{\mathbf{W}}_{s/2}, \hat{\mathbf{E}}_{s/2}\right)$ *of* $\left(\mathbf{W}_{s/2}, \mathbf{E}_{s/2}\right)$ *up to an ambiguity matrix, i.e., there exist a nonsingular matrix* $Q \in \mathbb{R}^{n\times n}$ *such that*

$$\hat{\mathbf{W}}_{s/2} = \mathbf{W}_{s/2}Q, \;\; \hat{\mathbf{E}}_{s/2} = Q^{-1}\mathbf{E}_{s/2}.$$

*Proof: First,* we will show that the optimal solution to (20) can yield the estimate of $\left(\mathbf{W}_{s/2}, \mathbf{E}_{s/2}\right)$ up to an ambiguity matrix.

From the matrix quantities in (18) and (19), we can see that structured matrices $\mathbf{O}_{2R+1,s/2}$ and $\mathbf{C}_{2R+1,s/2}$ are linearly parameterized by the block components of $\mathbf{W}_{s/2}$ and $\mathbf{E}_{s/2}$, respectively. By the result of Theorem 2, it can be derived that the optimal solution to (20) can yield the estimate of $\left(\mathbf{W}_{s/2}, \mathbf{E}_{s/2}\right)$ up to an ambiguity matrix.

*Second,* we will show that the optimal solution to (23) can yield the estimate of $\left(\mathbf{W}_{s/2}, \mathbf{E}_{s/2}\right)$ up to an ambiguity matrix.

By equation (22), the optimization problem (20) can be reformulated as

$$\min_{\mathbf{W},\mathbf{E}} \quad \|\mathbf{H}_{2R+1,2R+1} - \mathcal{H}(\mathbf{W}\mathbf{E})\|_F^2, \tag{24}$$

where the variables $\mathbf{W}$ and $\mathbf{E}$ have the same sizes of $\mathbf{W}_{s/2}$ and $\mathbf{E}_{s/2}$, respectively. It is noted that equation (24) is a parameterized form of the structured low-rank matrix factorization problem in (20).

By Theorem 2 and Lemma 5, the optimal solution $(\hat{\mathbf{W}}, \hat{\mathbf{E}})$ to (24) satisfies $\text{rank}\left(\hat{\mathbf{W}}\hat{\mathbf{E}}\right) = n$ and $\mathcal{H}(\hat{\mathbf{W}}\hat{\mathbf{E}}) = \mathbf{H}_{2R+1,2R+1}$. It can be derived that $\mathbf{X} = \hat{\mathbf{W}}\hat{\mathbf{E}}$ is an optimal solution to (23) and the criterion of (23) becomes zero.

Let $\hat{\mathbf{X}}$ be an optimal solution to (23). It should satisfy $\mathcal{H}(\hat{\mathbf{X}}) = \mathbf{H}_{2R+1,2R+1}$ and $\text{rank}\left(\hat{\mathbf{X}}\right) = n$. Let the SVD decomposition of $\hat{\mathbf{X}}$ be given by $\hat{\mathbf{X}} = \hat{U}\Sigma\hat{V}^T$, where $\hat{U}$ and $\hat{V}$ are constructed by $n$ orthogonal columns, and $\Sigma \in \mathbb{R}^{n\times n}$ is a diagonal matrix with positive diagonal entries. Then, $(\hat{\mathbf{W}}, \hat{\mathbf{E}}) = (\hat{U}, \Sigma\hat{V}^T)$ is an optimal solution to (24). Therefore, $(\hat{U}, \Sigma\hat{V}^T)$ is an estimate of $\left(\mathbf{W}_{s/2}, \mathbf{E}_{s/2}\right)$ up to an ambiguity matrix. The proof is then completed.

■

The rank-constrained optimization problem in (23) is non-convex and NP-hard. Following our previous work [27], the rank-constrained optimization problem (23) is recast into a difference-of-convex optimization problem which is then solved using the sequential convex programming method.

Let $\hat{\mathbf{X}}$ be an optimal solution to (23). The SVD decomposition of $\hat{\mathbf{X}}$ is given by

$$\hat{\mathbf{X}} = \begin{bmatrix} U_1 & U_2 \end{bmatrix} \begin{bmatrix} \Sigma_1 & \\ & \Sigma_2 \end{bmatrix} \begin{bmatrix} V_1^T \\ V_2^T \end{bmatrix}, \tag{25}$$

where $U_1$ and $V_1$ consists of $n$ orthogonal columns, and the diagonal matrix $\Sigma_1 \in \mathbb{R}^{n\times n}$ has larger diagonal entries than $\Sigma_2$. The estimates of $\mathbf{W}_{s/2}$ and $\mathbf{E}_{s/2}$ can then be obtained as follows:

$$\hat{\mathbf{W}}_{s/2} = U_1, \qquad \hat{\mathbf{E}}_{s/2} = \Sigma_1 V_1^T.$$

Since $\mathbf{O}_{2R+1,s/2}$ and $\mathbf{C}_{2R+1,s/2}$ are respectively affine in terms of $\mathbf{W}_{s/2}$ and $\mathbf{E}_{s/2}$, the variables $\mathbf{O}$ and $\mathbf{C}$ in (20) can be estimated from $\hat{\mathbf{W}}_{s/2}$ and $\hat{\mathbf{E}}_{s/2}$, respectively.

### *C. Determining the system matrices $\{A, A_l, A_r, B, C\}$*

In view of the theoretical result in Theorem 2, we assume that the obtained estimates $\hat{\mathbf{O}}_{2R+1,s/2}$ and $\hat{\mathbf{C}}_{2R+1,s/2}$ satisfy

$$\begin{aligned} \hat{\mathbf{O}}_{2R+1,s/2} &= \mathbf{O}_{2R+1,s/2}\mathbf{Q}, \\ \hat{\mathbf{C}}_{2R+1,s/2} &= \mathbf{Q}^{-1}\mathbf{C}_{2R+1,s/2}, \end{aligned} \tag{26}$$

where $\mathbf{Q} = I_{2R+1} \otimes Q$ with $Q \in \mathbb{R}^{n\times n}$ being nonsingular. Based on these estimates, we will address the identification of the system matrices $\{A, A_l, A_r, B, C\}$ up to a similarity transformation.

First, the shifting structure of the time-varying observability matrix $\mathbf{O}_{2R+1,s/2}$ will be explored. Denote

$$\mathbf{O}_{j,k_1:k_2} = \begin{bmatrix} \underline{C}_{j-2k_1} G_{j-2k_1} \cdots G_{j-2} \\ \underline{C}_{j-2(k_1+1)} G_{j-2(k_1+1)} \cdots G_{j-2} \\ \vdots \\ \underline{C}_{j-2k_2} G_{j-2k_2} \cdots G_{j-2} \end{bmatrix}$$

where $0 \le k_1 < k_2 \le s/2 - 2$ and $2k_2 \le j \le 2R+1$. The matrix $\mathbf{O}_{j,k_1:k_2}$ above is constructed by the block rows of $\mathbf{O}_{j,k}$ with block-row indices from $k_1$ to $k_2$. Then, the *structure-shifting property* of $\mathbf{O}_{2R+1,s/2}$ can be represented as

$$\mathbf{O}_{2R-1,0:s/2-2} G_{2R-1} = \mathbf{O}_{2R+1,1:s/2-1}, \tag{27}$$

where $\mathbf{O}_{2R-1,0:s/2-2}$ and $\mathbf{O}_{2R+1,1:s/2-1}$ are sub-matrices of $\mathbf{O}_{2R+1,s/2}$, and $G_{2R-1}$ is a block Toeplitz matrix defined in Definition 2. This is illustrated in the following example.

**Example 5.** *When $R = 2$ and $s = 6$, the structure-shifting property in* (27) *can be explicitly written as*

$$\left[\begin{array}{ccc} C & & \\ & C & \\ & & C \\ \hline CA_l & CA & CA_r \end{array}\right] \begin{bmatrix} A_l & A & A_r & & \\ & A_l & A & A_r & \\ & & A_l & A & A_r \end{bmatrix}$$
$$= \left[\begin{array}{ccccc} CA_l & CA & CA_r & & \\ & CA_l & CA & CA_r & \\ & & CA_l & CA & CA_r \\ \hline CA_l^2 & C(AA_l + A_lA) & * & C(AA_r + A_rA) & CA_r^2 \end{array}\right],$$

*where $*$ is used to represent the term $C(A_lA_r + A^2 + A_rA_l)$.*

Based on equation (27), we formulate the following structured least-squares optimization problem to identify the matrices $A_l, A, A_r$ based on the estimate $\hat{\mathbf{O}}_{2R+1,s/2}$:

$$\begin{aligned} \min_{G} \quad & \|\hat{\mathbf{O}}_{2R-1,0:s/2-2} G - \hat{\mathbf{O}}_{2R+1,1:s/2-1}\|_F^2 \\ s.t. \quad & G \in \mathcal{G}_{2R-1}, \end{aligned} \tag{28}$$

where $\mathcal{G}_{2R-1}$ denotes a set of matrices having the same structure as $G_{2R-1}$, as shown in Definition 2.

The optimal solution to (28) has properties shown in the following lemma.

**Lemma 6.** *Let $\hat{\mathbf{O}}_{2R+1,s/2}$ satisfy equation* (26)*. Assume that $\hat{\mathbf{O}}_{2R-1,0:s/2-2}$ has full column rank. Then, the optimal solution $\hat{G}$ to the optimization problem in* (28) *satisfies*

$$\hat{G} = \left(I_{2R-1} \otimes Q^{-1}\right) G_{2R-1} \left(I_{2R+1} \otimes Q\right), \tag{29}$$

*where $Q \in \mathbb{R}^{n\times n}$ is a nonsingular matrix.*

The above lemma can be derived straightforwardly based on equation (26) and the optimization formulation (28). The condition that the time-varying observability matrix $\hat{\mathbf{O}}_{2R-1,0:s/2-2}$ has full column rank is similar to condition 3) of Theorem 2. Lemma 6 implies that the matrices $A_l, A, A_r$ can be determined up to a similarity transformation, i.e.,

$$\hat{A}_l = Q^{-1}A_lQ, \;\; \hat{A} = Q^{-1}AQ, \;\; \hat{A}_r = Q^{-1}A_rQ.$$

In addition, according to equation (26), the estimates $\hat{C}$ and $\hat{B}$ can be extracted respectively from $\hat{\mathbf{O}}_{2R+1,s/2}$ and $\hat{\mathbf{C}}_{2R+1,s/2}$, satisfying that

$$\hat{C} = CQ, \;\; \hat{B} = Q^{-1}B.$$

To ease the reference, the proposed local network identification algorithm is summarized in Algorithm 1.

| **Algorithm 1**: Local identification for 1D distributed systems | |
|---|---|
| Step 1 | Construct a spatially stacked state-space model (3) and its temperally stacked equation (4) based on local observations; |
| Step 2 | Estimate $\mathcal{T}_s^{B_R}$ from the optimization problem (12); |
| Step 4 | Estimate $\mathbf{O}_{2R+1,s/2}$ and $\mathbf{C}_{2R+1,s/2}$ using the method described in Subsection V-B; |
| Step 5 | Extract the estimates of $C$ and $B$ from the estimates of $\mathbf{O}_{2R+1,s/2}$ and $\mathbf{C}_{2R+1,s/2}$, respectively; |
| Step 6 | Estimate $A_l, A, A_r$ by solving (28). |

## VI. Numerical simulation

In this section, numerical simulations are provided to demonstrate the effectiveness of the proposed identification method – Algorithm 1. In the simulation, the distributed system is constructed by connecting 40 identical subsystems in a line, and the identification for the 20-th subsystem is performed. The system matrices $(A, A_l, A_r, B, C)$ with $A, A_l, A_r \in \mathbb{R}^{3\times 3}$, $B \in \mathbb{R}^{3\times 2}$ and $C \in \mathbb{R}^{2\times 3}$ are randomly generated such that Assumption A.1 is satisfied and the 1D networked system is stable.

To construct an augmented state-space model in (3), we set $R = 5$ and $s = 8$. The system input and the measurement noise are generated as white noise sequences and the data length is set to 800.

To evaluate the identification performance against the noise effect, the criterion signal-to-noise ratio (SNR) is adopted, which is defined as

$$\text{SNR} = 10 \log\left(\frac{\text{var}(y_i(k) - e_i(k))}{\text{var}(e_i(k))}\right).$$

In the sequel, we shall carry out numerical simulations with the SNR ranging from 0 dB to 95 dB.

We use the criterion *impulse-response fitting* to evaluate the performance of the proposed identification method. The normalized fitting error of the impulse-response sequence $CA^iB$ is defined by

$$\frac{1}{T}\sum_{j=1}^{T}\frac{\sum_{i=0}^{10}\|CA^iB-\hat{C}_j\hat{A}_j^i\hat{B}_j\|_F}{\sum_{i=0}^{10}\|CA^iB\|_F},$$

where $T$ is the number of randomly generated networked systems and $\{\hat{C}_j,\hat{A}_j,\hat{B}_j\}$ are the estimates of $\{C,A,B\}$ of the $j$-th generated networked system, respectively. Similarly, we can also define the normalized fitting error for the impulse-response sequences $CA_l^iB$ and $CA_r^iB$.

Fig. 3 shows the identification performance of the proposed method against the SNR. The normalized fitting errors are calculated by averaging the results of 100 randomly generated networked systems, and the regularization parameter $\lambda$ in (8) is set to $\lambda=10^{-3}$. It can be observed from Fig. 3 that, using the proposed identification method, the normalized fitting error decreases along with the increase of the SNR. When the SNR is larger than 50 dB, the normalized fitting errors can be as small as $10^{-4}$.

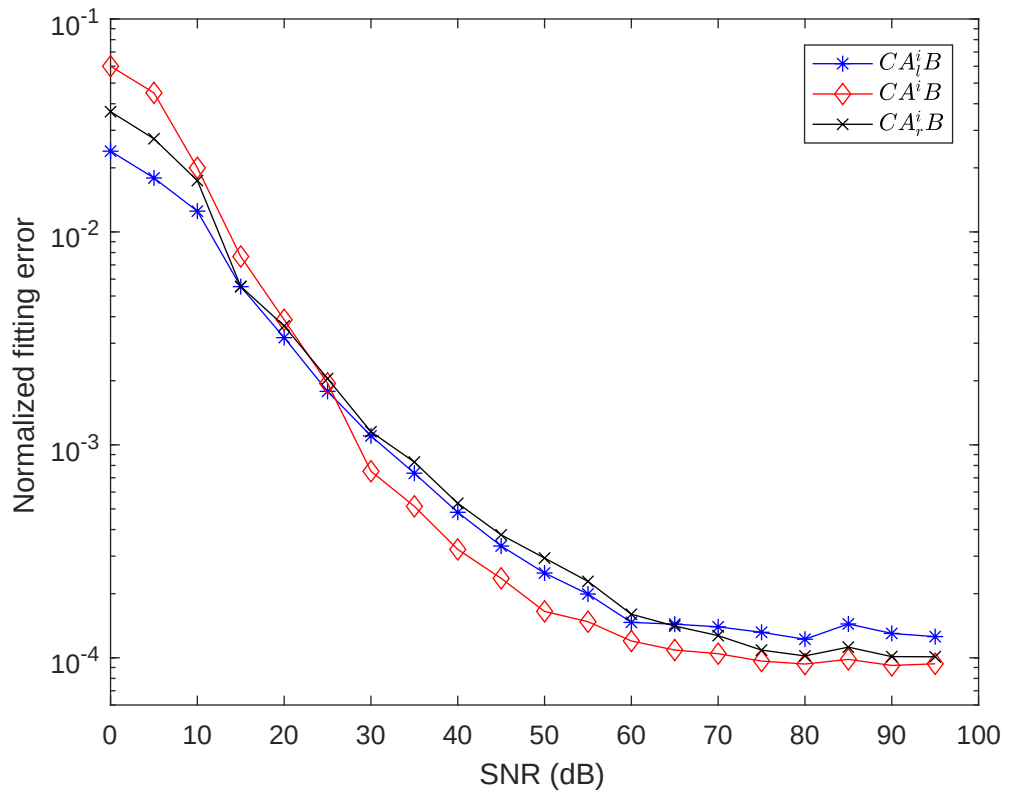

Fig. 3. Normalized fitting errors under different noise levels with $\lambda=10^{-3}$.

Fig. 4 shows the impulse-response fidelity of individual subsystems against the regularization parameter $\lambda$, where SNR is set to SNR=40 dB. The normalized fitting errors are computed by averaging the results of 100 randomly generated networked systems. We can observe from Fig. 4 that good identification performance can be achieved by choosing the regularization parameter $\lambda$ around $10^{-2}$.

## VII. Conclusion

The local identification of 1-D large-scale distributed systems has been studied. Compared with the classical system identification problems, the challenging point of the local system identification is that there are two unknown system inputs which are the states of its neighboring subsystems. By exploiting both the spatial and temporal structures of the distributed system, especially the two-layer Toeplitz structure of the Markov-parameter matrix, a low-rank optimization problem has been provided for identifying the Markov parameters of a local cluster of identical subsystems, where the associated optimal solution can yield (parts of) the true Markov parameters. Moreover, the system realization of a structured state-space model is formulated as a structured low-rank matrix factorization problem, showing that the system matrices can be determined up to a similarity transformation by enforcing the structure of the generalized observability/controllability matrix.

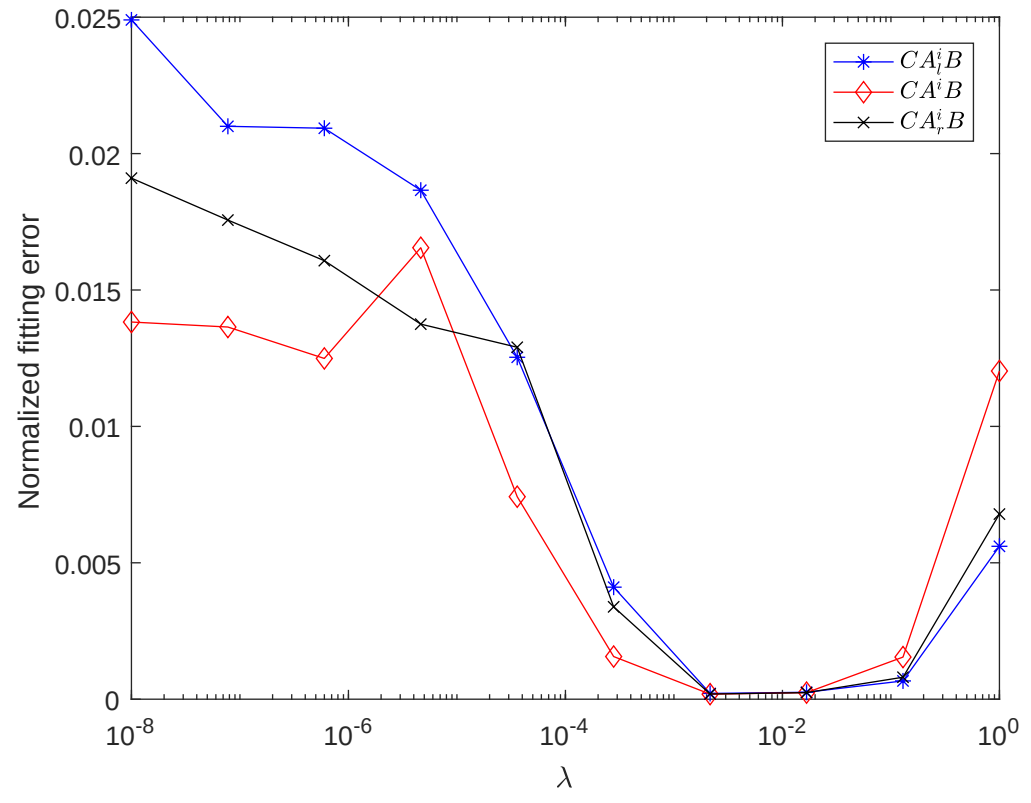

Fig. 4. Normalized fitting errors against the regularization parameter $\lambda$ with SNR=40 dB.

Although we only consider the identification of 1D homogeneous networked systems, it can be extended to 2D homogeneous networks using the same identification framework, namely exploiting both the spatial and temporal structures of the concerned distributed system. In our future work, the identification of large heterogeneous networks will be investigated.

## Appendix A
## Proof of Lemma 2

The proof of the lemma is divided into three steps.

*Step I*: we shall show that $\begin{bmatrix} x_h^i \\ V_{s,h}^i \\ U_{s,h}^i \end{bmatrix}$ has full row rank under the given assumptions in the lemma.

It can be derived from (2) that

$$\begin{bmatrix} x(k) & \cdots \\ x(k+1) & \cdots \\ \vdots & \cdots \\ x(k+s-1) & \cdots \end{bmatrix} = \begin{bmatrix} I & & & \\ \mathcal{A} & \mathcal{B} & & \\ \vdots & \vdots & \ddots & \\ \mathcal{A}^{s-1} & \mathcal{A}^{s-2}\mathcal{B} & \cdots & \mathcal{B} \end{bmatrix} \times \begin{bmatrix} x(k) & \cdots \\ u(k) & \cdots \\ \vdots & \cdots \\ u(k+s-2) & \cdots \end{bmatrix}. \quad (30)$$

Let
$\Xi=\begin{bmatrix} O_{(2R+1)n,(i-R-1)n} & I_{(2R+1)n} & O_{(2R+1)n,(N-i-R)n} \end{bmatrix}$,
$\Omega=\begin{bmatrix} O_{n,(i-R-2)n} & I_n & O_{n,(2R+1)n} & O_{n,n} & O_{n,(N-i-R-1)n} \\ O_{n,(i-R-2)n} & O_{n,n} & O_{n,(2R+1)n} & I_n & O_{n,(N-i-R-1)n} \end{bmatrix}$
and
$\Pi=\begin{bmatrix} O_{(2R+1)m,(i-R-1)m} & I_{(2R+1)m} & O_{(2R+1)m,(N-i-R)m} \end{bmatrix}$

be selection matrices such that

$$\begin{bmatrix} x_h^i \\ V_{s,h}^i \\ U_{s,h}^i \end{bmatrix} = \begin{bmatrix} \Xi & & & & \\ \Omega & & & & \\ \Omega\mathcal{A} & \Omega\mathcal{B} & & & \\ \vdots & \vdots & \ddots & & \\ \Omega\mathcal{A}^{s-1} & \Omega\mathcal{A}^{s-2}\mathcal{B} & \cdots & \Omega\mathcal{B} & 0 \\ 0 & \Pi & & & \\ \vdots & & \ddots & & \\ 0 & & & \Pi & \\ 0 & & & & \Pi \end{bmatrix} \times \begin{bmatrix} x(k) & \cdots \\ u(k) & \cdots \\ \vdots & \cdots \\ u(k+s-2) & \cdots \\ u(k+s-1) & \cdots \end{bmatrix}. \tag{31}$$

By Lemma 10.4 in [20], under the assumption that $u(k)$ is persistently exciting of order $nN+s$, it can be established that

$$\begin{bmatrix} x(k) & \cdots \\ u(k) & \cdots \\ \vdots & \cdots \\ u(k+s-1) & \cdots \end{bmatrix}$$

has full row rank. In addition, by explicitly unfolding the expressions of $\Omega\mathcal{A}^j\mathcal{B}$, it can be verified that, when $(A_l, B)$ and $(A_r, B)$ have full row rank and $R \geq s-1$, the coefficient matrix on the right-hand side of (31) has full row rank. For the sake of brevity, we only show this for the case with $s=3$. The associated coefficient matrix in (31) is shown in (32) with $*$ denoting unimportant entries for analyzing the concerned rank property. Under the condition that $(A_l, B)$ and $(A_r, B)$ have full row rank, it is easy to see that the sub-matrix stacked by the first five and last two block rows of (32) has full row rank. It can be further verified that the whole coefficient matrix in (32) has full row rank. This is a consequence of $\begin{bmatrix} A_l^2 & A_lB & B \end{bmatrix} = \begin{bmatrix} A_l & B \end{bmatrix} \begin{bmatrix} A_l & B & 0 \\ 0 & 0 & I \end{bmatrix}$ and $\begin{bmatrix} A_r^2 & A_rB & B \end{bmatrix} = \begin{bmatrix} A_r & B \end{bmatrix} \begin{bmatrix} A_r & B & 0 \\ 0 & 0 & I \end{bmatrix}$ with $\begin{bmatrix} A_l & B \end{bmatrix}$ and $\begin{bmatrix} A_r & B \end{bmatrix}$ having full row rank.

Now that the coefficient matrix in (31) has full row rank, Sylvesters' inequality shows that $\begin{bmatrix} x_h^i \\ V_{s,h}^i \\ U_{s,h}^i \end{bmatrix}$ has full row rank as well.

*Step II:* Let $\mathbf{T}_s^{B_R} \in \mathcal{T}$ be the true value of $\Theta_s^{B_R}$. Next, we shall show that $\mathbf{T}_s^{B_R}$ is an optimal solution to (9). Let $\Delta_s \in \mathcal{T}$. Under the full row rank property of $\begin{bmatrix} x_h^i \\ V_{s,h}^i \\ U_{s,h}^i \end{bmatrix}$, we can derive that

$$\begin{aligned} &\text{rank}\left[Y_{s,h}^i - \mathbf{T}_s^{B_R} U_{s,h}^i\right] \\ =&\text{rank}\left(\begin{bmatrix} \mathcal{O}_s & \mathcal{T}_s^{D_R} \end{bmatrix}\begin{bmatrix} x_h^i \\ V_{s,h}^i \end{bmatrix}\right) \\ =&\text{rank}\begin{bmatrix} \mathcal{O}_s & \mathcal{T}_s^{D_R} \end{bmatrix} \end{aligned}$$

and

$$\begin{aligned} &\text{rank}\left[Y_{s,h}^i - \left(\mathbf{T}_s^{B_R} - \Delta_s\right) U_{s,h}^i\right] \\ =&\text{rank}\left(\begin{bmatrix} \mathcal{O}_s & \mathcal{T}_s^{D_R} & \Delta_s \end{bmatrix}\begin{bmatrix} x_h^i \\ V_{s,h}^i \\ U_{s,h}^i \end{bmatrix}\right) \\ =&\text{rank}\begin{bmatrix} \mathcal{O}_s & \mathcal{T}_s^{D_R} & \Delta_s \end{bmatrix}. \end{aligned} \tag{33}$$

By the inequality

$$\text{rank}\begin{bmatrix} \mathcal{O}_s & \mathcal{T}_s^{D_R} \end{bmatrix} \leq \text{rank}\begin{bmatrix} \mathcal{O}_s & \mathcal{T}_s^{D_R} & \Delta_s \end{bmatrix}, \tag{34}$$

we can obtain that

$$\begin{aligned} &\text{rank}\left[Y_{s,h}^i - \mathbf{T}_s^{B_R} U_{s,h}^i\right] \\ &\quad \leq \text{rank}\left[Y_{s,h}^i - \left(\mathbf{T}_s^{B_R} - \Delta_s\right) U_{s,h}^i\right] \end{aligned}$$

for any $\Delta_s \in \mathcal{T}$. Thus, $\mathbf{T}_s^{B_R}$ is an optimal solution to (9).

*Step III:* In this step, we shall show that the optimal solution to (9) is non-unique. Let

$$\Delta_s = \mathcal{T}_s^{D_R}\mathcal{G}_s, \tag{35}$$

where $\mathcal{G}_s = I_s \otimes G_R$ with $G_R \in \mathbb{R}^{2n\times(2R+1)n}$. Denote $\mathbb{T}_s^{B_R} = \mathbf{T}_s^{B_R} - \Delta_s$. It is easy to verify that $\mathbb{T}_s^{B_R} \in \mathcal{T}$.

Since $\Delta_s = \mathcal{T}_s^{D_R}\mathcal{G}_s$, it can be established that

$$\begin{aligned} &\text{rank}\left([Y_{k,s,h}^i - \mathbb{T}_s^{B_R} U_{k,s,h}^i\right) \\ &= \text{rank}\begin{bmatrix} \mathcal{O}_s & \mathcal{T}_s^{D_R} & \Delta_s \end{bmatrix} \\ &= \text{rank}\left(\begin{bmatrix} \mathcal{O}_s & \mathcal{T}_s^{D_R} \end{bmatrix}\begin{bmatrix} I & 0 & 0 \\ 0 & I & \mathcal{G}_s \end{bmatrix}\right) \\ &= \text{rank}\begin{bmatrix} \mathcal{O}_s & \mathcal{T}_s^{D_R} \end{bmatrix} \\ &= \text{rank}\left([Y_{s,h}^i - \mathbf{T}_s^{B_R} U_{s,h}^i\right). \end{aligned} \tag{36}$$

It can then be observed from the above equation that both $\mathbb{T}_s^{B_R}$ and $\mathbf{T}_s^{B_R}$ are optimal solutions to (9); therefore, the optimal solution to (9) is non-unique.

## Appendix B
## Proof of Theorem 1

Following the proof of Lemma 2, the following matrix in (31) has full row rank:

$$\begin{bmatrix} x_h^i \\ V_{s,h}^i \\ U_{s,h}^i \end{bmatrix}.$$

Let $\mathcal{T}_f$ denote the set of the two-layer block Toeplitz matrices, defined in Subsection IV-B. Let $\mathbf{T}_s^{B_R} \in \mathcal{T}_f$ be the true value of the estimate of $\Theta_s^{B_R}$ in (14). Since the matrix set $\mathcal{T}_f$ belongs to the matrix set $\mathcal{T}$, the minimal value of the cost function in (9) is smaller than or equal to that in (14), i.e.

$$\begin{aligned} &\min_{\Theta_s^{B_R}\in\mathcal{T}} \text{rank}\left[Y_{s,h}^i - \Theta_s^{B_R} U_{s,h}^i\right] \\ &\quad \leq \min_{\Theta_s^{B_R}\in\mathcal{T}_f} \text{rank}\left[Y_{s,h}^i - \Theta_s^{B_R} U_{s,h}^i\right]. \end{aligned}$$

As shown in Lemma 2, the true value $\mathbf{T}_s^{B_R} \in \mathcal{T}_f$ is an optimal solution to (9). Then, it can be observed from the above inequality that $\mathbf{T}_s^{B_R} \in \mathcal{T}_f$ is also an optimal solution to

$$
\left[\begin{array}{ccc|ccc|ccc|ccc|c|ccc|ccc|c|ccc}
0 & 0 & 0 & I_n & * & I_n & 0 & 0 & 0 & & & & & & & & & & & & & & \\
\hline
0 & 0 & I_n & 0 & * & 0 & 0 & 0 & 0 & & & & & & & & & & & & & & \\
0 & 0 & 0 & 0 & * & 0 & I_n & 0 & 0 & & & & & & & & & & & & & & \\
\hline
0 & A_l & A & A_r & * & 0 & 0 & 0 & 0 & 0 & B & 0 & * & 0 & 0 & 0 & & & & & & & \\
0 & 0 & 0 & 0 & * & A_l & A & A_r & 0 & 0 & 0 & 0 & * & 0 & B & 0 & & & & & & & \\
\hline
A_l^2 & * & * & * & * & 0 & 0 & 0 & 0 & A_lB & AB & A_rB & * & 0 & 0 & 0 & 0 & B & 0 & * & 0 & 0 & 0 \\
0 & 0 & 0 & 0 & * & * & * & * & A_r^2 & 0 & 0 & 0 & * & A_lB & AB & A_rB & 0 & 0 & 0 & * & 0 & B & 0 \\
\hline
& & & & & & & & & & & & I_{(2R+1)m} & & & & & & & 0 & & & \\
\hline
& & & & & & & & & & & & 0 & & & & & & & I_{(2R+1)m} & & &
\end{array}\right] \tag{32}
$$

(14). As a consequence, the corresponding minimal value of the cost function of (14) is

$$\operatorname{rank}\left[\begin{array}{cc} \mathcal{O}_s & \mathcal{T}_s^{D_R} \end{array}\right].$$

Let $\mathbb{T}_s^{B_R} \in \mathcal{T}_f$ be such that $\Delta_s = \mathbf{T}_s^{B_R} - \mathbb{T}_s^{B_R}$ is a perturbation we seek of the true value such that it retains the minimal value of the cost function in (14). Using the full row rank condition of the matrix in (31), we obtain that

$$
\begin{aligned}
&\operatorname{rank}\left(Y_{s,h}^i - \mathbb{T}_s^{B_R} U_{s,h}^i\right) \\
&= \operatorname{rank}\left[\mathcal{O}_s x_h^i + \mathcal{T}_s^{D_R} V_{s,h}^i + \Delta_s U_{s,h}^i\right] \\
&= \operatorname{rank}\left[\begin{array}{ccc} \mathcal{O}_s & \mathcal{T}_s^{D_R} & \Delta_s \end{array}\right].
\end{aligned} \tag{37}
$$

Now, we seek that part of $\Delta_s \in \mathcal{T}_f$ that retains the minimal value of the cost function in (14) only when it is zero. For that reason, we permute the block rows of $\Delta_s$ compatible with the zero block-row pattern of the matrix $\mathcal{T}_s^{D_R}$.

*(a) Structure of $\mathcal{T}_s^{D_R}$:* The structure of $\mathcal{T}_s^{D_R}$ is determined by $\underline{C}_R \underline{A}_R^j \underline{D}_R$ for $j = 0, 1, \cdots, s-2$. By the definitions of $\underline{A}_R, \underline{C}_R$ and $\underline{D}_R$ in (3), we can find that $\underline{C}_R \underline{A}_R^j \underline{D}_R$ is a $(2R+1) \times 2$ block matrix. When $R \geq (s-1)$, the block rows of $\underline{C}_R \underline{A}_R^j \underline{D}_R$ indexed from $(j+2)$ to $(2R-j)$ are zero. For instance, $\underline{C}_R \underline{A}_R^j \underline{D}_R$ has the block structure as shown below with $*$ representing non-zero block entries:

$$
\left[\begin{array}{c|c}
* & \\
\vdots & 0 \\
* & \\
\hline
0 & 0 \\
\hline
 & * \\
0 & \vdots \\
 & *
\end{array}\right]. \tag{38}
$$

*(b) Permutation of $\Delta_s$:* Let $L_s$ denote the matrix that permutes all zero block rows of $\mathcal{T}_s^{D_R}$ to top such that

$$L_s \mathcal{T}_s^{D_R} = \left[\begin{array}{c} 0 \\ T^2 \end{array}\right], \tag{39}$$

where $T^2$ is a sub-matrix of $\mathcal{T}_s^{D_R}$ by removing the corresponding zero block rows, as illustrated in (38). It is noteworthy that the zero block rows of $\mathcal{T}_s^{D_R}$ correspond to the second-layer block Toeplitz part of $\mathcal{T}_s^{B_R}$, as shown in Lemma 3.

Then, the permutation matrix $L_s$ is applied to the matrices $\Delta_s$ and $\mathcal{O}_s$, yielding the following special structures

$$
\begin{aligned}
L_s \Delta_s &= \left[\begin{array}{c} \Delta^1 \\ \hline * \end{array}\right] \\
&= \left[\begin{array}{ccccc}
0 & & & & \\
\Delta_{0,0} & 0 & & & \\
\Delta_{1,0} & \Delta_{0,1} & 0 & & \\
\vdots & \ddots & \ddots & \ddots & \\
\Delta_{s-2,0} & \cdots & \Delta_{1,s-3} & \Delta_{0,s-2} & 0 \\
\hline
 & & * & &
\end{array}\right], \\
L_s \mathcal{O}_s &= \left[\begin{array}{c} \mathcal{O}_s^1 \\ \hline * \end{array}\right],
\end{aligned} \tag{40}
$$

such that $\mathcal{O}_s^1 = \mathbf{O}_{2R+1,s}$ and $\Delta_{j,l}$, for $j = 0, 1, \cdots, s-2$ and $l = 0, \cdots, s-2-j$, is a $[2(R-j-l)-1] \times [2R+1]$ block Toeplitz matrix in the form

$$
\Delta_{j,l} = \left[\begin{array}{cccccccc}
0 & \delta_{j,-j} & \delta_{j,1-j} & \cdots & \delta_{j,j} & 0 & & \\
 & \ddots & \ddots & \ddots & \ddots & \ddots & \ddots & \\
 & & 0 & \delta_{j,-j} & \delta_{j,1-j} & \cdots & \delta_{j,j} & 0
\end{array}\right],
$$

with $\{\delta_{j,k}\}_{k=-j}^{j}$ being block entries of appropriate sizes. Note that $\Delta_{j,l}$ for $l \geq 1$ is a sub-matrix of $\Delta_{j,0}$, which is constructed by stacking the block rows indexed from $(l+1)$ to $(2R-1-l)$.

Now, we are ready to show for what values of $\Delta_{j,l}$ the following rank constraint holds

$$\operatorname{rank}\left[\begin{array}{ccc} \mathcal{O}_s & \mathcal{T}_s^{D_R} & \Delta_s \end{array}\right] = \operatorname{rank}\left[\begin{array}{cc} \mathcal{O}_s & \mathcal{T}_s^{D_R} \end{array}\right]. \tag{41}$$

It can be derived from the above equality that

$$L_s \left[\begin{array}{cc} \mathcal{O}_s & \mathcal{T}_s^{D_R} \end{array}\right] \left[\begin{array}{c} Q_1 \\ Q_2 \end{array}\right] = L_s \Delta_s, \tag{42}$$

or

$$\left[\begin{array}{c|c} \mathbf{O}_{2R+1,s} & 0 \\ \hline * & T^2 \end{array}\right] \left[\begin{array}{c} Q_1 \\ Q_2 \end{array}\right] = \left[\begin{array}{c} \Delta^1 \\ \hline * \end{array}\right], \tag{43}$$

where $Q_1$ and $Q_2$ are coefficient matrices of appropriate sizes, and $*$ denotes non-zero block entries caused by the block-row permutation. So we focus on the top part of (43) and rewrite it as

$$\Delta^1 = \mathbf{O}_{2R+1,s} \left[\begin{array}{cccc} Q_{1,0} & Q_{1,1} & \cdots & Q_{1,s-1} \end{array}\right], \tag{44}$$

where $\{Q_{1,j}\}_{j=0}^{s-1}$ are sub-matrices of $Q_1$ satisfying that $Q_1 = [Q_{1,0} \;\; Q_{1,1} \; \cdots \; Q_{1,s-1}]$.

By considering the last block column of equation (44), under the assumption that $\mathbf{O}_{2R+1,s-1}$ has full column rank, we can obtain that $Q_{1,s-1} = 0$. Similarly, by considering the second to the last block column, it can be derived that $Q_{1,s-2} = 0$ and $\Delta_{0,s-2} = 0$, further $\Delta_{0,j} = 0$ for $j = 0, 1, \cdots, s-3$. Next, for the third to the last block column, since $\Delta_{0,s-3} = 0$ and $\mathbf{O}_{2R+1,s-1}$ has full column rank, we can obtain that $Q_{1,s-3} = 0$ and $\Delta_{1,j} = 0$ for $j = 0, \cdots, s-3$. By repeating this procedure, it can be established that equation (44) holds only when $\Delta^1 = 0$ and $Q_{1,j} = 0$ for $j = 0, 1, \cdots, s-1$. In other words, the matrix $\Delta_s$ with a nontrivial sub-matrix $\Delta^1$ cannot be linearly represented by $\begin{bmatrix} \mathcal{O}_s & \mathcal{T}_s^{D_R} \end{bmatrix}$.

Since the block rows of $\Delta^1$ correspond to the second-layer Toeplitz part of $\Delta_s$, we can see that the rank of $\begin{bmatrix} \mathcal{O}_s & \mathcal{T}_s^{D_R} & \Delta_s \end{bmatrix}$ is strictly larger than that of $\begin{bmatrix} \mathcal{O}_s & \mathcal{T}_s^{D_R} \end{bmatrix}$ as long as the second-layer Toeplitz part of $\Delta_s$ is non-zero. Thus, the proof is completed.

## Appendix C
## Proof of Theorem 2

Assume that the matrix pair $\{\hat{\mathbf{O}}_{2R+1,s/2}, \hat{\mathbf{C}}_{2R+1,s/2}\}$ is one of the optimal solution pairs to (20), it can be established that $\hat{\mathbf{O}}_{2R+1,s/2}$ and $\mathbf{O}_{2R+1,s/2}$ has the same structure and the same column space, which can be algebraically represented as

$$\hat{\mathbf{O}}_{2R+1,s/2} = \mathbf{O}_{2R+1,s/2}\mathbf{Q}, \tag{45}$$

where $\mathbf{Q} \in \mathbb{R}^{(2R+1)n \times (2R+1)n}$ is a $(2R+1, 2R+1)$ block matrix which is nonsingular. In the sequel, we denote $Q_{i,j}$ the $(i,j)$-th block entry of $\mathbf{Q}$. In order to prove the theorem, it is sufficient to prove that the estimate $\hat{\mathbf{W}}_{s/2}$ of $\mathbf{W}_{s/2}$ satisfies that

$$\hat{\mathbf{W}}_{s/2} = \mathbf{W}_{s/2}Q, \tag{46}$$

where $Q \in \mathbb{R}^{n \times n}$ is a nonsingular matrix and $\mathbf{W}_{s/2}$ is defined in (19). In the sequel, we denote by $\hat{W}_{i,j}$, for $j = -i, \cdots, i$, the estimate of $W_{i,j}$ which is a block component of $\mathbf{W}_{s/2}$.

To illustrate the structure of equation (45), it is expanded into equation (47) by setting $R = 2$ and $s = 4$.

The following proof will be divided into three steps.

*Step 1.* We will show that, under the assumption that $\mathbf{O}_{2R,s/2}$ has full column rank, the following relations can be derived from equation (45):

$$\begin{aligned}
i = 0: \quad & \hat{W}_{i,-i} = W_{i,-i}Q_{1,1}, \\
& W_{i,-i}Q_{1,j} = 0 \text{ for } j = 2, \cdots, s-1; \\
i = 1: \quad & \hat{W}_{i,-i} = W_{i,-i}Q_{1,1}, \\
& \hat{W}_{i,1-i} = W_{i,-i}Q_{1,2} + W_{i,1-i}Q_{1,1}, \\
& \hat{W}_{i,2-i} = W_{i,-i}Q_{1,3} + W_{i,1-i}Q_{1,2} + W_{i,2-i}Q_{1,1}, \\
& W_{i,-i}Q_{1,j+2} + W_{i,1-i}Q_{1,j+1} + W_{i,2-i}Q_{1,j} = 0 \\
& \qquad\qquad \text{for } j = 2, \cdots, s-3; \\
& \cdots \\
i = s/2 - 1: \quad & \hat{W}_{i,-i} = W_{i,-i}Q_{1,1}, \\
& \cdots \\
& \hat{W}_{i,s-2-i} = W_{i,-i}Q_{1,s-1} + \cdots + W_{i,s-2-i}Q_{1,1}.
\end{aligned} \tag{48}$$

Let's consider the equation corresponding to the first block-column of (45). Since $\mathbf{O}_{2R,s/2}$ has full column rank, it can be derived that

$$\begin{aligned}
& Q_{i,1} = 0 \text{ for } i = 2, 3, \cdots, 2R+1; \\
& \hat{W}_{i,-i} = W_{i,-i}Q_{1,1} \text{ for } i = 0, 1, \cdots, s/2 - 1.
\end{aligned} \tag{49}$$

To illustrate this, the first block-column equation of (47) can be equivalently written as follows:

$$\left[\begin{array}{c} \hat{C} \\ \hat{CA_l} \\ \hline 0 \\ 0 \\ 0 \\ 0 \\ 0 \\ 0 \end{array}\right] = \left[\begin{array}{c|cccc} C & 0 & 0 & 0 & 0 \\ CA_l & CA & CA_r & 0 & 0 \\ \hline 0 & C & 0 & 0 & 0 \\ 0 & 0 & C & 0 & 0 \\ 0 & 0 & 0 & C & 0 \\ 0 & 0 & 0 & 0 & C \\ 0 & CA_l & CA & CA_r & 0 \\ 0 & 0 & CA_l & CA & CA_r \end{array}\right] \begin{bmatrix} Q_{1,1} \\ Q_{2,1} \\ Q_{3,1} \\ Q_{4,1} \\ Q_{5,1} \end{bmatrix}.$$

It is obvious that, since $\mathbf{O}_{4,2}$ (or $\mathbf{O}_{2R,s/2}$) has full column rank, it has

$$\begin{aligned}
& Q_{2,1} = Q_{3,1} = Q_{4,1} = Q_{5,1} = 0, \\
& \hat{C} = CQ_{1,1}, \quad \hat{CA_l} = CA_lQ_{1,1}.
\end{aligned} \tag{50}$$

Next, by substituting the relations in (49) into the equation corresponding to the second block-column equation of (45) and under the assumption that $\mathbf{O}_{2R,s/2}$ has full column rank, we can obtain that

$$\begin{aligned}
& Q_{i,2} = 0 \text{ for } i = 3, \cdots, 2R+1; \qquad CQ_{1,2} = 0; \\
& \hat{W}_{i,1-i} = W_{i,-i}Q_{1,2} + W_{i,1-i}Q_{1,1} \text{ for } i = 1, \cdots, s/2 - 1.
\end{aligned} \tag{51}$$

To illustrate this, by using the relations in (50), the second block-column equation of (47) can be equivalently written as

$$\left[\begin{array}{c} \hat{CA} \\ 0 \\ \hline CQ_{1,1} \\ 0 \\ 0 \\ 0 \\ CA_lQ_{1,1} \\ 0 \end{array}\right] = \left[\begin{array}{c|cccc} CA_l & CA & CA_r & & \\ C & & & & \\ \hline & C & & & \\ & & C & & \\ & & & C & \\ & & & & C \\ & CA_l & CA & CA_r & \\ & & CA_l & CA & CA_r \end{array}\right] \begin{bmatrix} Q_{1,2} \\ Q_{2,2} \\ Q_{3,2} \\ Q_{4,2} \\ Q_{5,2} \end{bmatrix}. \tag{52}$$

We can observe that the bottom subpart of the vector on the left-hand side equals the product of the first block column of $\mathbf{O}_{4,2}$ and $Q_{1,1}$. By the assumption that $\mathbf{O}_{4,2}$ (or $\mathbf{O}_{2R,s/2}$) has full column rank, we can derive that

$$\begin{aligned}
& Q_{3,2} = Q_{4,2} = Q_{5,2} = 0; \quad CQ_{1,2} = 0; \quad Q_{2,2} = Q_{1,1}; \\
& \hat{CA} = CA_lQ_{1,2} + CAQ_{1,1}.
\end{aligned} \tag{53}$$

By iteratively considering from the first to the $(s-1)$-th block-column equation of (45) using the same strategy, we can derive the relations in equation (48).

*Step 2.* Under the assumption that $\mathbf{O}_{2R,s/2}$ has full column rank, the following relations can be derived from equation (45)

$$\left[\begin{array}{ccccc} \hat{C} & & & & \\ & \hat{C} & & & \\ & & \hat{C} & & \\ & & & \hat{C} & \\ & & & & \hat{C} \\ \hline \hat{CA_l} & \hat{CA} & \hat{CA_r} & & \\ & \hat{CA_l} & \hat{CA} & \hat{CA_r} & \\ & & \hat{CA_l} & \hat{CA} & \hat{CA_r} \end{array}\right] = \left[\begin{array}{ccccc} C & & & & \\ & C & & & \\ & & C & & \\ & & & C & \\ & & & & C \\ \hline CA_l & CA & CA_r & & \\ & CA_l & CA & CA_r & \\ & & CA_l & CA & CA_r \end{array}\right] \begin{bmatrix} Q_{1,1} & Q_{1,2} & Q_{1,3} & Q_{1,4} & Q_{1,5} \\ Q_{2,1} & Q_{2,2} & Q_{2,3} & Q_{2,4} & Q_{2,5} \\ Q_{3,1} & Q_{3,2} & Q_{3,3} & Q_{3,4} & Q_{3,5} \\ Q_{4,1} & Q_{4,2} & Q_{4,3} & Q_{4,4} & Q_{4,5} \\ Q_{5,1} & Q_{5,2} & Q_{5,3} & Q_{5,4} & Q_{5,5} \end{bmatrix} \tag{47}$$

as well:

$$\begin{aligned} i=0: \quad & \hat{W}_{i,i} = W_{i,i}Q_{2R+1,2R+1}, \\ & W_{i,i}Q_{2R+1,j} = 0 \quad \text{for } j = 2R+3-s, \cdots, 2R; \\ i=1: \quad & \hat{W}_{i,i} = W_{i,i}Q_{2R+1,2R+1}, \\ & \hat{W}_{i,i-1} = W_{i,i}Q_{2R+1,2R} + W_{i,i-1}Q_{2R+1,2R+1}, \\ & \hat{W}_{i,i-2} = W_{i,i}Q_{2R+1,2R-1} + W_{i,i-1}Q_{2R+1,2R} \\ & \qquad + W_{i,i-2}Q_{2R+1,2R+1}, \\ & W_{i,i}Q_{2R+1,j} + W_{i,i-1}Q_{2R+1,j+1} \\ & \qquad\qquad + W_{i,i-2}Q_{2R+1,j+2} = 0 \\ & \qquad\qquad \text{for } j = 2R+3-s, \cdots, 2R-2; \\ & \cdots \\ i = s/2-1: \quad & \hat{W}_{i,i} = W_{i,i}Q_{2R+1,2R+1}, \\ & \cdots \\ & \hat{W}_{i,i-s+2} = W_{i,i}Q_{2R+1,2R+3-s} + \cdots \\ & \qquad\qquad + W_{i,i-s+2}Q_{2R+1,2R+1}. \end{aligned} \tag{54}$$

The relations in (54) are derived using the same strategy in *Step 1* by iteratively considering from the $(2R+1)$-th to the $(2R+3-s)$-th block-column equation of (45) in a reverse order.

*Step 3.* We will show $\hat{\mathbf{W}}_{s/2} = \mathbf{W}_{s/2}Q_{1,1}$ using the relations in (48) and (54).

Denote by $\hat{\mathbf{O}}_{2s-3,s/2}(:,s-1)$ the $(s-1)$-th block column of the estimate $\hat{\mathbf{O}}_{2s-3,s/2}$, which is a sub-matrix of $\hat{\mathbf{O}}_{2R+1,s/2}$. The relations in (48) can be compactly written as:

$$\hat{\mathbf{O}}_{2s-3,s/2}(:,s-1) = \mathbf{O}_{2s-3,s/2} \begin{bmatrix} Q_{1,s-1} \\ \vdots \\ Q_{1,1} \\ 0 \\ \vdots \\ 0 \end{bmatrix}, \tag{55}$$

while those in (54) can be compactly written as:

$$\hat{\mathbf{O}}_{2s-3,s/2}(:,s-1) = \mathbf{O}_{2s-3,s/2} \begin{bmatrix} 0 \\ \vdots \\ 0 \\ Q_{2R+1,2R+1} \\ \vdots \\ Q_{2R+1,2R+3-s} \end{bmatrix} \tag{56}$$

To illustrate this using the specific example shown in (47), the equations (55)-(56) can be explicitly written as

$$\left[\begin{array}{c} 0 \\ 0 \\ \hat{C} \\ 0 \\ 0 \\ \hline \hat{CA_r} \\ \hat{CA} \\ \hat{CA_l} \end{array}\right] = \mathbf{O}_{5,2} \begin{bmatrix} Q_{1,3} \\ Q_{1,2} \\ Q_{1,1} \\ 0 \\ 0 \end{bmatrix}, \qquad \left[\begin{array}{c} 0 \\ 0 \\ \hat{C} \\ 0 \\ 0 \\ \hline \hat{CA_r} \\ \hat{CA} \\ \hat{CA_l} \end{array}\right] = \mathbf{O}_{5,2} \begin{bmatrix} 0 \\ 0 \\ Q_{5,5} \\ Q_{5,4} \\ Q_{5,3} \end{bmatrix}. \tag{57}$$

In view of equations (55) and (56), we can obtain that

$$\mathbf{O}_{2s-3,s/2} \begin{bmatrix} Q_{1,s-1} \\ \vdots \\ Q_{1,2} \\ Q_{1,1} \\ 0 \\ \vdots \\ 0 \end{bmatrix} = \mathbf{O}_{2s-3,s/2} \begin{bmatrix} 0 \\ \vdots \\ 0 \\ Q_{2R+1,2R+1} \\ Q_{2R+1,2R} \\ \vdots \\ Q_{2R+1,2R+3-s} \end{bmatrix}$$

By assumption 3) of the theorem, the matrix $\mathbf{O}_{2s-3,s/2}$ has full column rank. As a result, it can be established from the above equation that

$$\begin{aligned} & Q_{1,1} = Q_{2R+1,2R+1}; \quad Q_{1,2} = \cdots = Q_{1,s-1} = 0; \\ & Q_{2R+1,2R+3-s} = \cdots = Q_{2R+1,2R} = 0. \end{aligned} \tag{58}$$

Since the block vector $\hat{\mathbf{O}}_{2s-3,s/2}(:,s-1)$ contains all the block components of $\hat{\mathbf{W}}_{s/2}$, we can obtain that

$$\hat{\mathbf{W}}_{s/2} = \mathbf{W}_{s/2}Q_{1,1}.$$

Therefore, the proof is completed.